\documentclass[12pt]{article}

\usepackage{macros}

\usepackage[margin = .95in]{geometry}

\pdfoutput=1

\usepackage{amsmath,amssymb,color}

\newcommand{\pkg}[1]{{\bf #1}}
\newcommand{\code}[1]{{\tt #1}}

\def\x{\mathbf{x}}
\def\y{\mathbf{y}}
\def\w{\mathbf{w}}
\def\r{\mathbf{r}}
\def\Rp{{R}}
\def\mypkg{\pkg{FKSUM}}
\def\R{\mathbb{R}}
\def\W{\mathbf{W}}
\def\Z{\mathbf{Z}}
\def\X{\mathbf{X}}
\def\p{\mathbf{p}}
\def\C{\mathcal{C}}
\def\ssig{\pmb{\Sigma}}
\def\one{\mathbf{1}}

\def\om{\pmb{\omega}}

\title{Fast Kernel Smoothing in {R} with Applications to Projection Pursuit}

\begin{document}

\author{\name David P.\ Hofmeyr \hfill
{\small \textmd{ Department of Statistics and Actuarial Science}}\\
\textcolor{white}{.}\hfill {\small \textmd{Stellenbosch University}}\\
\textcolor{white}{.}\hfill {\small \textmd{7600, South Africa}}
}

\maketitle

\begin{abstract}%
  This paper introduces the {R} package \pkg{FKSUM}, which offers fast and exact evaluation of univariate kernel smoothers. The main kernel computations are implemented in {C++}, and are wrapped in simple, intuitive and versatile {R} functions. The fast kernel computations are based on recursive expressions involving the order statistics, which allows for exact evaluation of kernel smoothers at all sample points in log-linear time. In addition to general purpose kernel smoothing functions, the package offers purpose built and ready-to-use implementations of popular kernel-type estimators. On top of these basic smoothing problems, this paper focuses on projection pursuit problems in which the projection index is based on kernel-type estimators of functionals of the projected density.
\end{abstract}

\begin{keywords}
kernel smoothing, non-parametric, density estimation, regression, projection pursuit, independent component analysis, {R}
\end{keywords}



\section{Introduction} \label{sec:intro}

Kernels offer an extremely flexible way of estimating (usually) smooth functions non-parametrically. At the essence of kernel smoothing, and indeed many non-parametric methods, is the simple concept of a local average around a point, $x$; that is, a weighted average of some observable quantities, those of which closest to $x$ being given the highest weights. Suppose our objective is the estimation of some structure (e.g., a function), which we cannot measure directly, and we instead make (indirect) observations subject to random error. If we can assume that these observations offer (close to) unbiased measurements of the function of interest, then in an ideal sampling scenario we would be able to make multiple such {\em noisy} measurements at each point of interest, so that highly efficient estimation can be achieved by taking the averages of these. In practice, however, we seldom have such control over how we sample observations, or there may be cost constraints which severely limit such direct approaches. Instead, we rely on the very simple observation that if the function of interest is continuous, then it will not change too substantially over a small region. Therefore, all observations which are in a neighbourhood of a point of interest should also provide reasonable information about the target. Averaging our observations, but placing almost all weight on those points in a neighbourhood of the target, is therefore very well motivated. 

Kernels provide an intuitive means for achieving such local averaging. In the most simple context (although in some contexts even the following may be relaxed), a kernel is simply a non-negative function which vanishes quickly as the magnitude of its argument increases. Kernels can be used as weighting functions if applied to the pairwise distances between points, so that this vanishing tendency ensures that weights associated with large distances between points are very small. Any weighted average arising from kernel weights will therefore only apply large weights to those points which are relatively near to the point of interest, i.e., in its neighbourhood. 
Furthermore, a simple rescaling of the function's domain, with a so-called {\em bandwidth}, allows one to control how quickly this vanishing occurs, and hence how large is this neighbourhood.

At the essence of kernel methods in statistics, then, lies the evaluation of sums of the form
\begin{align}\label{eq:ksum}
    S(x|\x,\om) := \sum_{j=1}^n K\left(\frac{x_j-x}{h}\right)\omega_j,
\end{align}
where $K(\cdot)$ is the kernel function, $x$ is a point of interest, $\x = (x_1,..., x_n)$ is a vector observed sample points and $\om = (\omega_1, ..., \omega_n)$ is for now an arbitrary vector of coefficients associated with the observations. These could be the noisy measurements of the structure/function being estimated, but as we will encounter in the remainder, may represent a variety of options. The parameter $h$ is the bandwidth, and it should be clear that if $h$ is relatively small then it will increase the magnitude of the arguments in $K(\cdot)$, so that the relative sizes of the associated weights will more heavily emphasise points close to $x$. The bandwidth is in a more general context referred to as a smoothing parameter, in that large values of $h$ lead to closer to uniform weights, and hence the total sum, $S(x|\x,\om)$, will vary less for different points, $x$ (i.e., will represent a smoother function of $x$). Arguably the most important areas of research in the context of kernel-type estimation is in the appropriate selection of the bandwidth, $h$, and in the design of efficient algorithms for evaluating $S(\cdot|\x,\om)$ for a large collection of evaluation points. In fact, as we will encounter, it is frequently necessary to compute these sums for all of the observations, $\x$, themselves. Na{\"i}ve evaluation of all such sums has computational complexity which is quadratic in $n$, which is prohibitive for even moderate sized problems. Some kernels with bounded support (those which take the value zero outside some compact set), as well as the Laplace kernel, allow so-called {\em fast sum updating}~\citep{langrene2019fast,fan1994fast,chen2006fast}, which means that these sums can be computed recursively, leading to log-linear computational complexity (the log factor arising from the fact that the observations and evaluation points must be sorted). A similar recursive approach was recently discussed for the class of kernels given by the product of a polynomial and the Laplace kernel~\citep{hofmeyr2019TPAMI}. Applications of the fast sum updating approach to products was also mentioned by~\cite{langrene2019fast}, although details are not given. The bounded support kernels tend to enjoy high {\em efficiency}, which relates to them inducing relatively low asymptotic mean integrated squared error (AMISE) when used for estimation. A limitation of these approaches is that they can result in all weights at a point being exactly zero. This may cause problems when using the estimated functions for prediction on some test data which include points outside of the range of the observations. They can also lead to less stable cross-validation for objectives such as maximum pseudo-likelihood, for the same reason. Other methods which are used for moderate-to-large sized problems rely on approximations, with popular examples being the fast Gauss and Fourier transforms~\citep[FGT,FFT]{YangDGD2003, Silverman1982}, and binning~\citep{ScottS1985, HallW1994}.

In this paper we will discuss the {R}~\citep{R} package \pkg{FKSUM}, which is available through the Comprehensive {R} Archive Network (CRAN). The package provides an implementation of the method described by~\cite{hofmeyr2019TPAMI} for fast and exact computation of sums as in Eq.~(\ref{eq:ksum}). The main kernel computations are implemented in {C++}, and accessed through {R} using the \pkg{Rcpp} package~\citep{Rcpp}. Because of their popularity in non-parametric statistics, there are numerous libraries which provide implementations of simple kernel smoothing methods. Popular examples include \pkg{KernSmooth}~\citep{kernsmooth}, \pkg{ks}~\citep{ks}, and \pkg{sm}~\citep{sm}, all of which are available through CRAN. As far as we are aware, however, no existing {R} libraries offer exact evaluation of kernel smoothers at all sample points in faster than quadratic running time. In addition to the basic univariate kernel estimators, \mypkg~offers implementations of multiple projection pursuit methods, including independent component analysis~\citep[ICA]{HyvarinenO2000}; projection pursuit regression~\citep[PPR]{friedman1981projection}; and minimum density hyperplane estimation for clustering~\citep[MDPP]{pavlidis2016minimum}. Multiple {R} libraries offer a variety of ICA models, including \pkg{fastICA}~\citep{CRANfastICA}, \pkg{PearsonICA}~\citep{CRANPearsonICA}, \pkg{ProDenICA}~\citep{CRANProDenICA} and \pkg{JADE}~\citep{CRANJADE}. A common and principled ICA objective is to minimise the mutual information in the estimated components, which is equivalent to minimising their individual differential entropies. Of the existing {R} implementations of which we are aware, \pkg{ProDenICA} is the only one which optimises a direct estimate of this objective. The most popular alternative used by other implementations is to replace this objective with a surrogate which measures departure from Gaussianity via a so-called {\em contrast function}. The motivation for this is that for any fixed variance the Gaussian distribution has the maximum differential entropy among all continuous distributions. PPR is implemented in {R}'s base \pkg{stats} package. In addition, the \pkg{gsg} package~\citep{gsgCRAN} provides generalisations to binary and Poisson responses. The implementation in \mypkg~provides functionality for the use of an arbitrary differentiable loss function, and so also offers considerable customisation. Finally, MDPP is implemented, along with other cluster motivated projection pursuit models, in the package \pkg{PPCI}~\citep{hofmeyr2019RJ}. The only other packages to combine projection pursuit with clustering explicitly, as far as we are aware, are \pkg{ProjectionBasedClustering}~\citep{ProjectionBasedClusteringCRAN} and \pkg{Pursuit}~\citep{PursuitCRAN}. The latter of these includes a range of projection pursuit models, including multiple exploratory methods, which may be seen as alternatives to ICA in certain contexts. 

This paper has two main objectives. The first is to provide the reader with a basic understanding of the methods implemented in the package, as well as the know-how for their use. The second is to provide more advanced readers with the tools to implement their own methods, or existing methods based on kernel smoothing which lie beyond the scope of the package. We provide explicit details for implementing the basic problems of kernel density estimation and regression, as well as instructions for how to estimate the bandwidth using cross-validation. We then also provide a very detailed description of the implementation of projection pursuit regression using the general purpose functions provided in the package. This example should give the more advanced reader coverage of the majority of challenges which they are likely to encounter in implementing their own projection pursuit methods, or methods outside the scope of the package.

\subsection{Getting started}

The \mypkg~package can be installed and loaded from within the \Rp~console with the commands
\begin{verbatim}
> install.packages('FKSUM')
> library(FKSUM)
\end{verbatim}
A brief introduction to the basic functionality of the package may then be accessed with the command
\begin{verbatim}
> help(FKSUM)
\end{verbatim}
The purpose built implementations offered in the package are kernel density estimation (\code{fk\_density}), kernel regression (\code{fk\_regression}), independent component analysis (\code{fk\_ICA}), projection pursuit regression (\code{fk\_ppr}) and minimum density hyperplanes (\code{fk\_mdh}). Details for the use of any function can be obtained from within the {R} console with the command \code{help(<function name>)}, e.g., \code{help(fk\_ICA)}.\\
\\
The remainder of the paper is organised as follows. In Section~\ref{sec:fksum} we introduce the use of the general purpose function for performing efficient and exact kernel smoothing. We go on to provide two simple but instructive examples of its use in the practical and important problems of density estimation and regression. In Section~\ref{sec:pp} we give a more detailed introduction to the general projection pursuit problem, before discussing the models and implementations in the package. In Section~\ref{sec:ppr} we provide a detailed discussion, with reference to the implementation of projection pursuit regression, which we hope will provide the reader with sufficient know-how for implementing their own projection pursuit methods which require efficient kernel evaluations. Finally, we give some concluding remarks.

\section[Fast Kernel Computations with FKSUM]{Fast Kernel Computations with \pkg{FKSUM}}\label{sec:fksum}

\begin{figure}
    \centering
    \includegraphics[width = \textwidth]{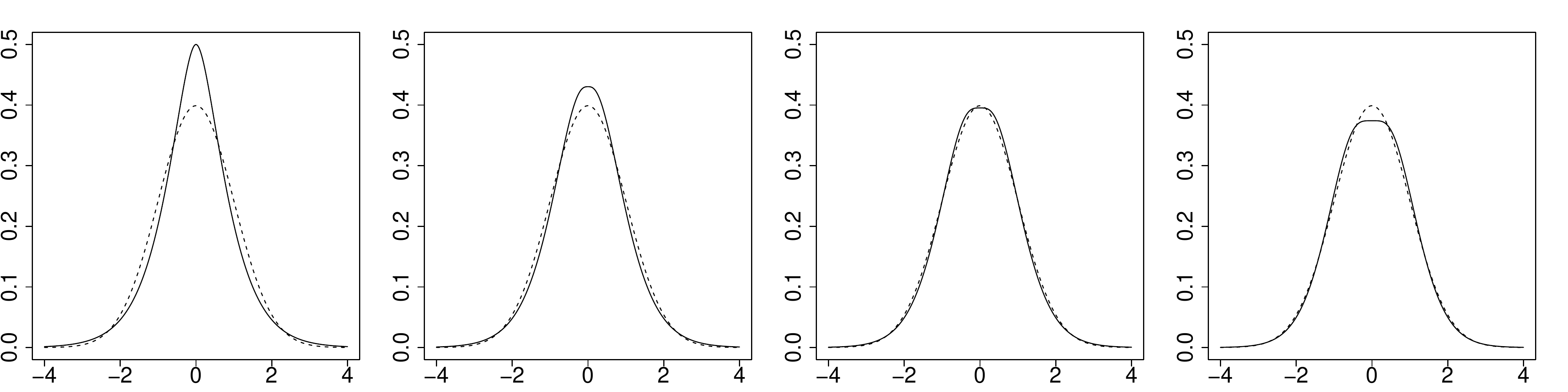}
    \caption{Smooth kernels of increasing order from one (left) to four (right). In addition the Gaussian kernel is shown (- - - -) for comparison. The order one kernel, in the leftmost plot, is the default used in the package.}
    \label{fig:smooth_kernels}
\end{figure}

In this section we introduce the general approach to perform kernel smoothing using the \pkg{FKSUM} package. Illustrations using the general purpose function \code{fk\_sum()}, which provides exact evaluation of sums of the form in Eq.~(\ref{eq:ksum}), will be provided. The function runs in $\mathcal{O}(n\log(n)+m\log(m))$ time for $n$ sample and $m$ evaluation points. The implementation is based on the method of~\cite{hofmeyr2019TPAMI}, which uses kernels which can be expressed in the form
\begin{align}\label{eq:polyexp}
    K(x) = \sum_{k=0}^\alpha \beta_k |x|^k \exp(-|x|),
\end{align}
for parameters $\beta_i>0, i = 0, ..., \alpha$. This is simply a product of an arbitrary polynomial in $|x|$ with positive coefficients and the Laplace kernel $K(x) \propto e^{-|x|}$. The value $\alpha$ is referred to as the {\em order} of the kernel, as it represents the order of the polynomial component (i.e., the highest exponent). The default kernel used in the package is an order one kernel with $\beta_0 = \beta_1 = \frac{1}{4}$. This is the simplest kernel of this form with two continuous derivatives. To visualise the shape of kernels of the type in Eq.~(\ref{eq:polyexp}), the package provides the function \code{plot\_kernel(beta, ...)}. The function requires only a single argument, which is the vector of coefficients $(\beta_0, ..., \beta_\alpha)$. In addition any graphical arguments accepted by \Rp's base \code{plot()} function are accepted. The visualisation produced by the function is scaled so that the kernel represents a probability density function of a random variable with unit variance. The normalising constant can be determined using the function \code{norm\_const\_K(beta)}. That is, if \code{beta} is a vector of positive coefficients, then the kernel with coefficients \code{beta/norm\_const\_K(beta)} has unit integral. In addition \code{var\_K(beta)} computes the variance of the random variable having as density the kernel with coefficients \code{beta/norm\_const\_K(beta)}. This standardisation of scale allows for far simpler visual comparison of the different kernels available in this class. \cite{hofmeyr2019TPAMI} discusses a sub-class of smooth kernels (a relatively high number of continuous derivatives at zero), for which $\beta_k \propto \frac{1}{k!}, k = 0, ..., \alpha$. The first four of these are shown in Figure~\ref{fig:smooth_kernels}, with the popular Gaussian kernel, $K(x) = \frac{1}{\sqrt{2\pi}}e^{-x^2/2}$, shown for comparison. The following code snippet can be used to reproduce this figure.
\begin{verbatim}
> par(mfrow = c(1, 4), mar = rep(2, 4))
>
> for(k in 1:4){
    plot_kernel(1 / factorial(0:k), ylim = c(0, .5))
    lines(seq(-4, 4, length = 500), dnorm(seq(-4, 4, length = 500)), lty = 2)
  }
\end{verbatim}
As can be seen in the figure, the order three kernel is visually similar to the Gaussian kernel. However, it is the order four kernel which has the closest efficiency to that of the Gaussian, for the purpose of density estimation. We have found that the default (order one) kernel is a very useful general purpose kernel, and have not often found reason to deviate from this choice. 
%

Now, as discussed in the introduction, at the essence of kernel-type estimators is sums of the form in Eq.~(\ref{eq:ksum}). Collections of such sums can be used for extremely flexible estimation of density functions, regression functions and spatial fields. However, of potentially greater importance, in some applications, than the estimation of a function, is the estimation of its derivative. Examples include when the function values themselves are not of great consequence, but the structure of the function, in terms of its stationary points, is the focus of the analysis. Furthermore, in the context of projection pursuit, when the objective function is based on a kernel-type estimator of the distribution of the projected data, it is necessary to compute sums of kernel derivatives in order to evaluate the gradient of the objective during optimisation. The \mypkg~package therefore also offers functionality for evaluating such sums exactly and efficiently.
Formally, the function \code{fk\_sum()} may be used to evaluate the collection of sums
\begin{align}\label{eq:ksum_dksum}
    \sum_{i=1}^n K\left(\frac{x_i - \tilde x_j}{h}\right)\omega_i \mbox{ and } \sum_{i=1}^n K'\left(\frac{x_i - \tilde x_j}{h}\right)\omega_i, \mbox{ for } j = 1, ..., m,
\end{align}
where $(x_1, ..., x_n)$ is a vector of univariate sample points, and $(\tilde x_1, ..., \tilde x_m)$ is a vector of evaluation points. The function takes the following arguments:
\begin{align*}
    \mbox{\code{x}}:& \mbox{ vector of sample points } (x_1, ..., x_n).\\
    \mbox{\code{omega}}:& \mbox{ vector of coefficients } (\omega_1, ..., \omega_n).\\
    \mbox{\code{h}}:& \mbox{ numeric bandwidth. Must be positive, i.e., \code{h > 0}.}\\
    \mbox{\code{x\_eval}}:& \mbox{ (optional) vector of evaluation points } (\tilde x_1, ..., \tilde x_m).\\
    & \mbox{ The default is \code{x\_eval = x}.}\\
    \mbox{\code{beta}}:& \mbox{ (optional) vector of kernel coefficients.}\\
    & \mbox{ The default is \code{beta = c(0.25, 0.25)}, corresponding to the smooth order 1 kernel.}\\
    \mbox{\code{nbin}}:& \mbox{ (optional) integer number of bins if binned estimator is to be used.}\\
    & \mbox{ If omitted then exact evaluation is performed.}\\
    \mbox{\code{type}}:& \mbox{ (optional) one of \code{"ksum"}, \code{"dksum"} and \code{"both"}. If \code{"ksum"} or \code{"dksum"}}\\
    & \mbox{ then the first or second set of sums in~(\ref{eq:ksum_dksum}), respectively, is returned. If \code{"both"}}\\
    & \mbox{ then the matrix \code{cbind(ksum, dksum)} is returned. The default is \code{type = "ksum"}.}
\end{align*}



\subsection[Simple Applications of fksum]{Simple Applications of \code{fk\_sum()}}

Before moving onto the main applications covered in the paper in the following section, we introduce the reader to the use of the \code{fk\_sum()} through a few examples. We consider two very simple examples which, although also implemented within purpose-built functions in the package, we believe provide straightforward but instructive applications of the function.

\paragraph{Example: Density estimation}

\begin{figure}
    \centering
    \includegraphics[height = 5cm, width = 10cm]{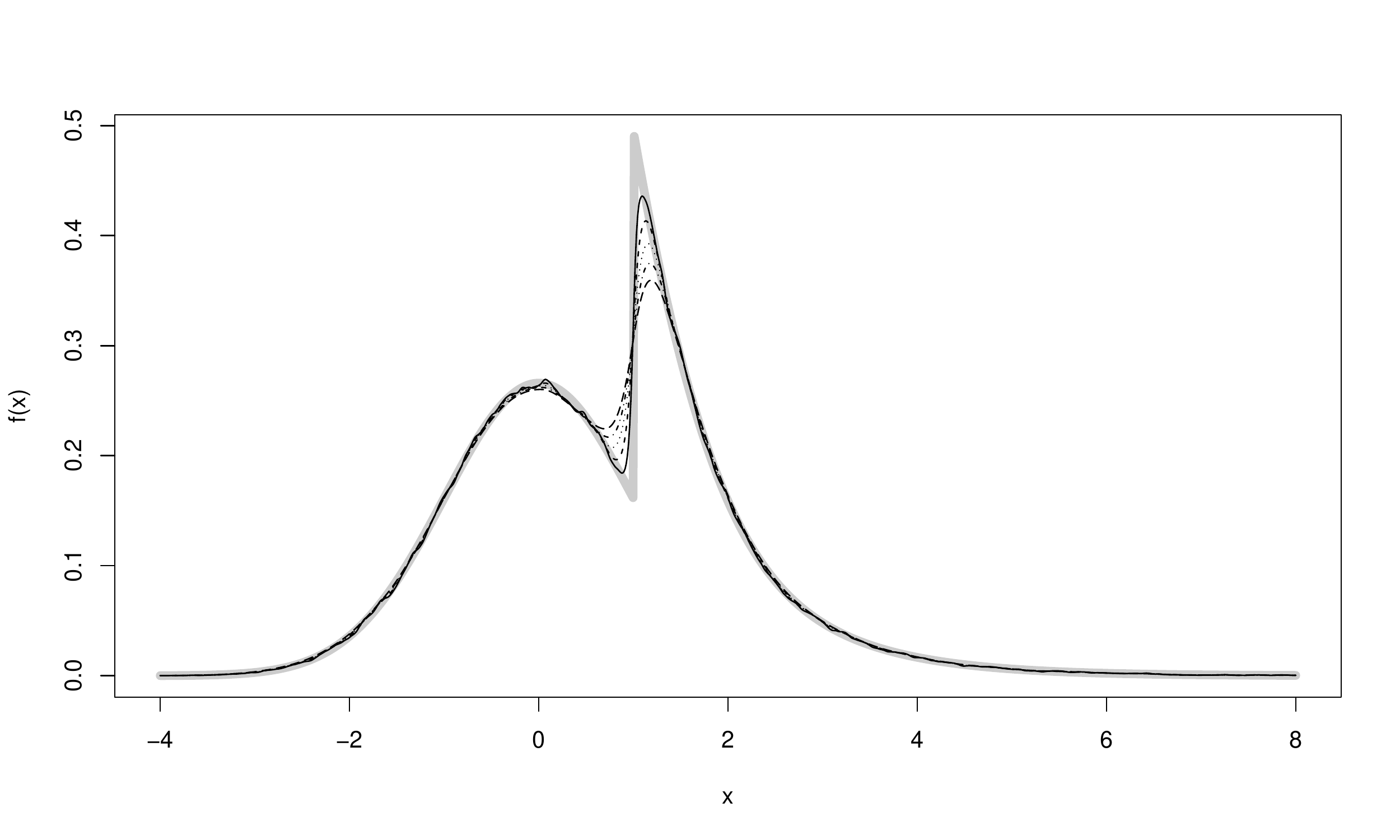}
    \includegraphics[height = 5cm, width = 4cm]{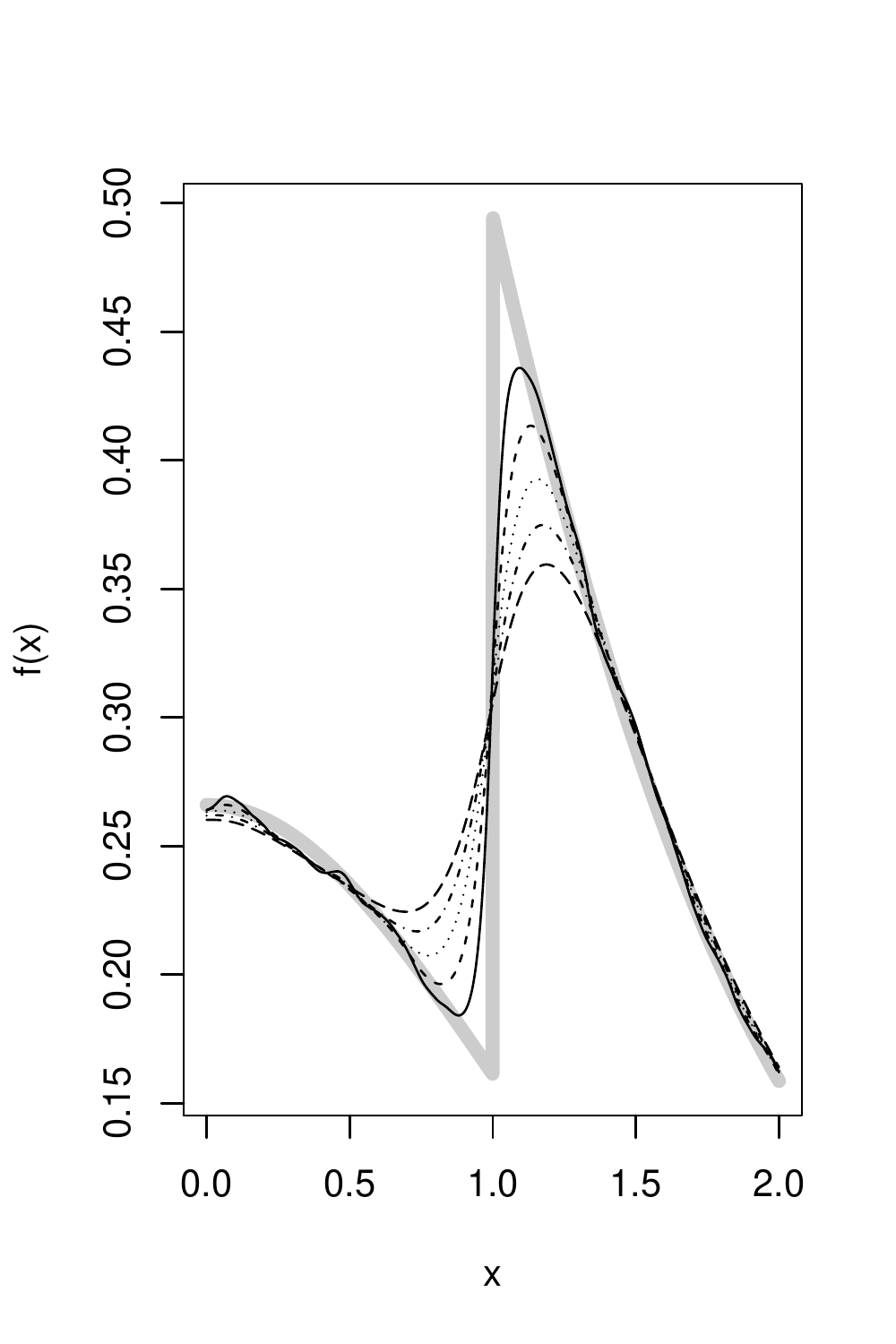}
    \caption{Bimodal mixture density. Full density (left) and zoomed section where differences are greatest (right). True density shown with thick grey line. Kernel estimates with varying bandwidths are shown with ------ ($h = 0.025$), - - - - ($h = 0.04375$), $\cdots\cdots$ ($h = 0.0625$), -$\cdot$-$\cdot$-$\cdot$- ($h = 0.08125$), --~--~--~($h = 0.1$)}
    \label{fig:density1}
\end{figure}

The simplest examples of Eq.~(\ref{eq:ksum}) are when the values of $\omega_i, i = 1, ..., n$, are all equal. The classic kernel density estimate is of this type. Specifically, the estimate of the underlying density, $f(\cdot)$, at a point $x$, is given by
\begin{align}\label{eq:kde}
    \hat f(x) = \frac{1}{nh}\sum_{j=1}^n K\left(\frac{x_j-x}{h}\right),
\end{align}
where the sample points $\{x_1, ..., x_n\}$ are assumed to be i.i.d. with density $f(\cdot)$.
In this example we will consider the density $f(x) = \frac{2}{3\sqrt{2\pi}}e^{-x^2/2} + \frac{1}{3}e^{-(x-1)}I(x>1)$, where $I(\cdot)$ is the indicator function. This is a bimodal mixture of a single Gaussian and a single exponential component. First we sample 150 000 points from this mixture,
\begin{verbatim}
> set.seed(1)  
> n <- 150000
> num_Gauss <- rbinom(1, n, 2 / 3)
> x <- c(rnorm(num_Gauss), rexp(n - num_Gauss) + 1)
\end{verbatim}
We now estimate the density for a range of bandwidth values, and plot the results, along with the true density. For illustrative purposes, we use the function \code{fk\_sum()} with the argument \code{omega} given by the constant vector with value $\frac{1}{nh}$, however the \pkg{FKSUM} package also includes the function \code{fk\_density()}. For more details, use the command \code{help(fk\_density)}.
\begin{verbatim}
> hs <- seq(.025, .1, length = 5)
> xeval <- seq(-4, 8, length = 1000)
> ftrue <- 2 / 3 * dnorm(xeval) + 1 / 3 * dexp(xeval - 1)
> plot(xeval, ftrue, lwd = 6, col = rgb(.8, .8, .8), xlab = "x",
        ylab = "f(x)", type = "l")
>
> for(i in 1:5) lines(xeval, fk_sum(x, rep(1 / hs[i] / n, n), hs[i],
        x_eval = xeval), lty = i)
\end{verbatim}
The results are shown in Figure~\ref{fig:density1}. The main differences between the estimates using different bandwidths lie in the region between 0 and 2, where the density function takes on its local extrema. 
%
The smaller bandwidths capture these most accurately, while larger bandwidths smooth over these extrema. Less obvious is that the smaller bandwidths lead to slightly less satisfying estimation at the mode located at zero, as they are less smooth, described by some, rather non-technically, as ``wiggly''. 
%

Selection of the bandwidth remains one of the challenges in kernel-type non-parametrics which gains a lot of attention. A very popular heuristic for density estimation, known informally as Silverman's rule of thumb, is based on the asymptotic mean integrated squared error (AMISE) minimiser under the assumption that the underlying distribution is Gaussian~\citep{Silverman1986}. In this case the bandwidth is given by
\begin{align*}
 h_{Silverman} = \left(\frac{8\sqrt{\pi}||K||^2}{3\sigma_K^4n}\right)^{1/5}\hat \sigma,   
\end{align*}
where $||K||^2 = \int_\R K(x)^2 dx, \sigma_K^2 = \int_\R x^2K(x) dx$, $n$ is the sample size and $\hat \sigma$ is an estimate of the standard deviation of the underlying random variable.
Many readers will likely be most familiar with this heuristic in the context of kernel density estimation using the Gaussian kernel, for which $||K||^2 = \frac{1}{2\sqrt{\pi}}$ and $\sigma_K^2 = 1$, and hence $h = \left(\frac{4}{3n}\right)^{1/5}\hat \sigma$. 
Both $||K||^2$ and $\sigma_K^2$ can be computed using the package, with the functions \code{roughness\_K()} and \code{var\_K()}, respectively. Each takes only a single argument, the vector of kernel coefficients, which will be normalised within the operation of the functions to ensure the kernel represents a density function. For the simulated data we therefore use the code,
\begin{verbatim}
> beta <- c(0.25, 0.25)
> const <- 8 * sqrt(pi) / 3
> h_sil <- (const * roughness_K(beta) / var_K(beta)^2 / n)^.2 * sd(x)
> h_sil
[1] 0.06841978 
\end{verbatim}
where we see that the selected value is close to the middle of the values used previously.
This approach is known to over-smooth densities with very sharp features, like the one shown in Figure~\ref{fig:density1}, and so sometimes a simple scaling is adopted. For example, the default bandwidth used by the R function \code{density()} is  $\approx 0.85 h_{Silverman}$.

The advantage of heuristics like this is that they require negligible computation. It should be clear, however, from the fact that their only dependence on the actual distribution is through the estimate $\hat \sigma$, that they may not be appropriate for all densities. A highly principled and general approach for selecting tuning parameters is cross-validation. In particular, the {\em leave-one-out} estimate of a function $f(\cdot)$, evaluated at one of the observations, say $x_i$, denoted $\hat f_{-i}(x_i)$, is simply the estimate of $f(x_i)$ determined from the sample $\{x_1, ..., x_{i-1}, x_{i+1}, ..., x_n\}$. That is, the estimate based on the points in the sample {\em excluding $x_i$}. For example, in the case of kernel density estimation we have $\hat f_{-i}(x_i) = \frac{1}{(n-1)h}\sum_{j\not = i}K\left(\frac{x_j - x_i}{h}\right)$. Notice that leave-one-out versions of kernel sums can easily be obtained by observing that
\begin{align*}
    \sum_{j\not = i} K\left(\frac{x_j-x_i}{h}\right)\omega_j = \sum_{j=1}^n K\left(\frac{x_j-x_i}{h}\right)\omega_j
    - K(0)\omega_i = 
    S(x_i|\x, \om) - \beta_0\omega_i.
\end{align*}
To compute the leave-one-out sums using \mypkg, we can therefore simply use the code \code{fk\_sum(x, omega, h, beta) - beta[1] * omega}. In the context of density estimation, the estimated (or pseduo-) likelihood, or its logarithm, may be used as an objective with respect to which cross-validation is applied. Practically, therefore, we would like to maximise the log-pseudo-likelihood (or equivalently minimise the negative likelihood) from the leave-one-out estimates as follows. First we define a function which evaluates this estimated log-likelihood for bandwidth \code{h}, data \code{x} and kernel coefficients \code{beta}.  Note that for small bandwidth values, it is possible that the leave-one-out estimates of the likelihood will be numerically evaluated to be zero. We therefore buffer these values at a small positive value to make the optimisation run more stably.
\begin{verbatim}
> loo_ml <- function(h, x, beta){
    n <- length(x)
    hf <- fk_sum(x, rep(1 / (n - 1) / h, n), h, beta = beta)
    hf_loo <- hf  - beta[1] / (n - 1) / h
    hf_loo[hf_loo < 1e-20] <- 1e-20
    -sum(log(hf_loo))
 }
\end{verbatim}
Next we apply the function \code{optimise()} from the base package \pkg{stats} to obtain the corresponding bandwidth value. The function's first and second arguments are the name of the function to be optimised, and the interval over which optimisation is to be performed. The following arguments are additional arguments accepted by the function being optimised. We provide the sorted data \code{sort(x)} so that the sorting needs to only be performed once, rather than in every call to \code{fk\_sum()}, which occurs inside the \code{loo\_ml()} function.
\begin{verbatim}
> h_ml <- optimise(loo_ml, sd(x) / length(x)^.2 * c(1 / 20, 5),
    sort(x), beta)$minimum
> h_ml
[1] 0.01526787
\end{verbatim}
Using complete information from the distribution of the sample means that this approach obtains a much smaller bandwidth value, which is better capable of capturing the local extrema and varying curvature of the underlying density. We can now plot the resulting estimates from these two bandwidth values, as before.

\begin{figure}
    \centering
    \includegraphics[height = 5cm, width = 10cm]{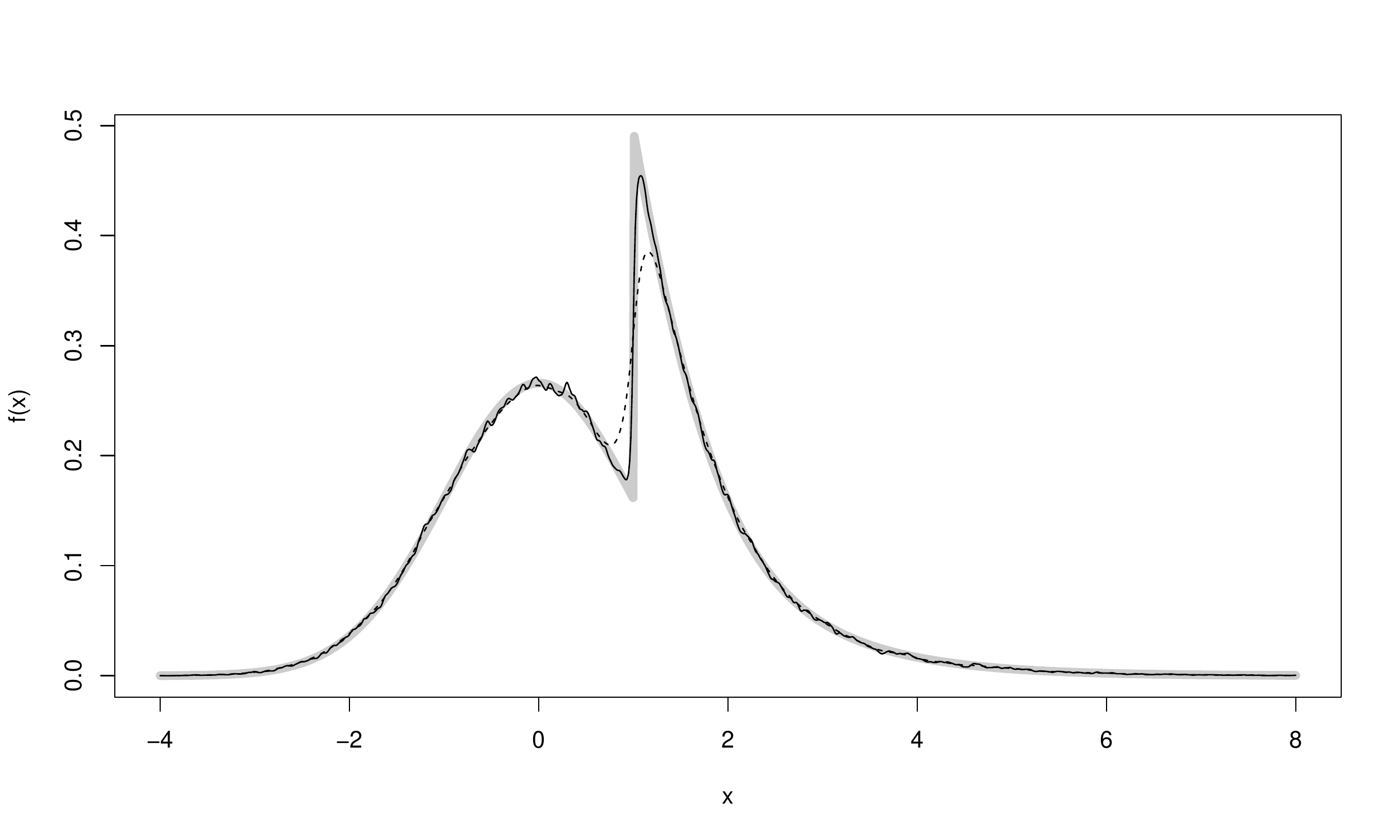}
    \includegraphics[height = 5cm, width = 4cm]{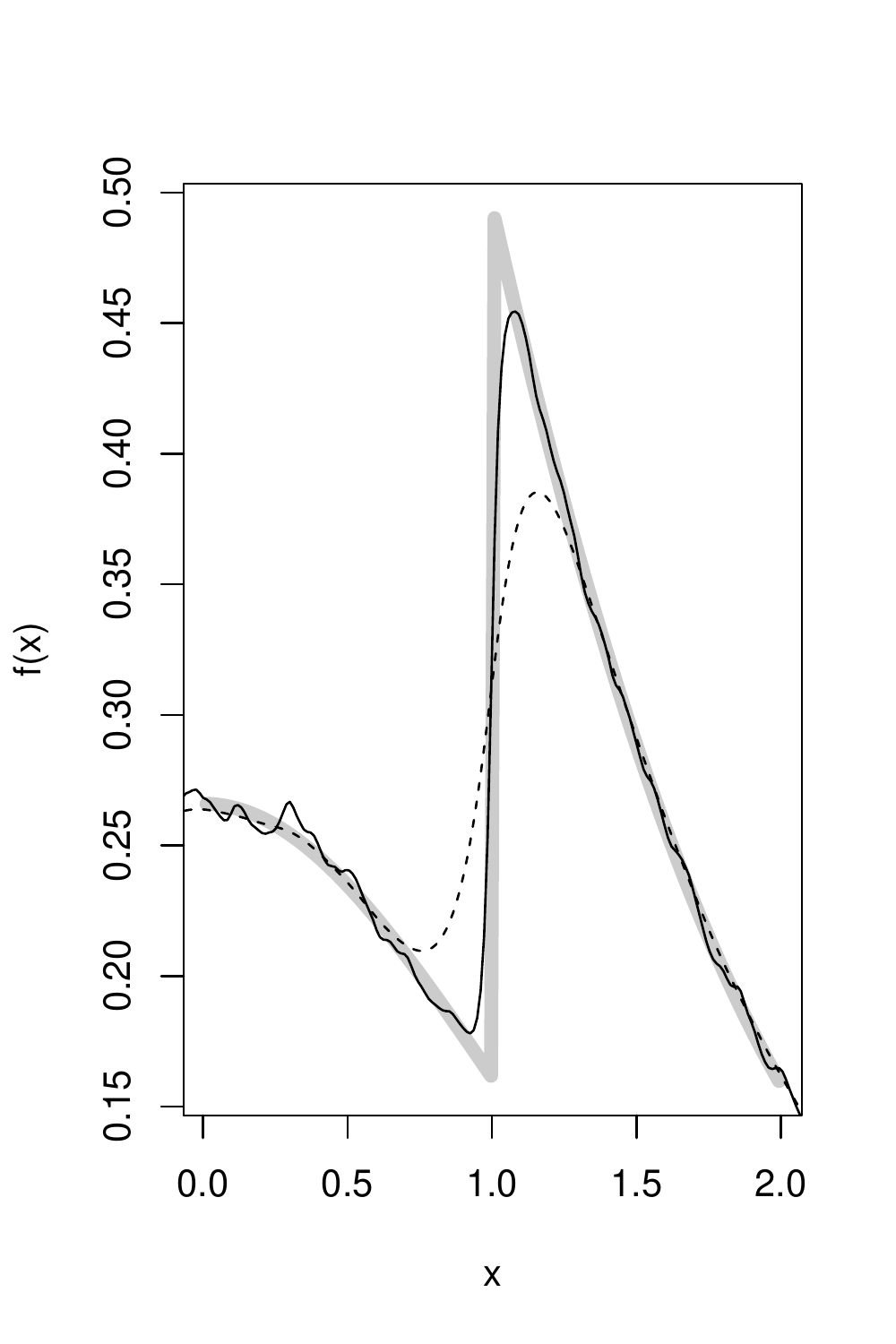}
    \caption{Bimodal mixture density. Full density (left) and zoomed section where differences are greatest (right). True density shown with thick grey line. Kernel estimates are shown with ------ ($h_{ML}$), and - - - - ($h_{Silverman}$)}
    \label{fig:density2}
\end{figure}

\begin{verbatim}
> plot(xeval, ftrue, lwd = 6, col = rgb(.8, .8, .8), xlab = "x",
        ylab = "f(x)", type = "l")
> lines(xeval, fk_sum(x, rep(1 / h_sil / n, n), h_sil,
        x_eval = xeval), lty = 2)
> lines(xeval, fk_sum(x, rep(1 / h_ml / n, n), h_ml, x_eval = xeval))
\end{verbatim}
The plots are shown in Figure~\ref{fig:density2} in the same manner as previously. We can see that the maximum pseudo-likelihood bandwidth gives a good estimation of the extrema of the function, but is considerably more wiggly than the estimate using Silverman's rule of thumb.

\paragraph{Example: Kernel regression}

In the standard formulation of the regression problem, the mean of a response variable, $Y$, is related to one or more covariates, $X_1, ..., X_d$, through some function $f(\cdot)$, which needs to be estimated. That is,
\begin{align}
Y = f(X_1, ..., X_d) + \epsilon,
\end{align}
where $\epsilon$ is a zero mean random variable, referred to as a {\em residual}, which may or may not be dependent on $X_1, ..., X_d$.
Here we assume that a single covariate, $X$, is available, and we observe a sample of pairs $\{(x_1, y_1), ..., (x_n, y_n)\}$, assumed to be independent realisations from the joint distribution of $X$ and $Y$. We will consider the simplest kernel-type estimator of the regression function, $f(\cdot)$, named after its two independent originators, the Nadaraya-Watson estimator~\citep{nadaraya, watson}. The estimator is based on the very simple reasoning that to estimate $f(x) = E[Y|X=x]$ we can take a weighted average of the observed response values, $y_1, ..., y_n$, which emphasises those for which the corresponding $x_i$'s are close to $x$.
Using kernels to provide these weights, we have
\begin{align}\label{eq:NWreg}
    \hat f(x) = \frac{\sum_{i=1}^n K\left(\frac{x_i-x}{h}\right)y_i}{\sum_{i=1}^n K\left(\frac{x_i-x}{h}\right)}.
\end{align}
As before, we sample a set of observations and plot the kernel based estimates for a range of bandwidth values. We generate data in which the underlying regression function is given by $f(x) = 3\sin(2x) + 10(x-5)I(x>5)$, where $I(\cdot)$ is again the indicator function. This is a sine function with a kink at the point $x=5$, above which a steep linear component is added. The residual distribution is equal to that of $T + (G-1)\left((X-5)^2+3\right)$, where $T\sim t_3$ and $G\sim$ Gamma$(2,2)$.
%
We sample 2000 pairs where $\frac{1}{10} X \sim $ Beta$(2,2)$ and the distribution of $Y|X$ is described above.
\begin{verbatim}
> set.seed(1)
> n <- 2000
> x <- rbeta(n, 2, 2) * 10
> fx <- 3 * sin(2 * x) + 10 * (x > 5) * (x - 5)
> y <- fx + rt(n, 3) + (rgamma(n, 2, 2) - 1) * ((x - 5)^2 + 3)
>
> hs <- seq(.05, .25, length = 5)
> xeval <- seq(0, 10, length = 1000)
> plot(x, y, col = rgb(.2, .2, .2, .2), pch = 16, xlab = "x", ylab = "f(x)")
> lines(xeval, 3 * sin(2 * xeval) + 10 * (xeval > 5) * (xeval - 5),
        col = 2, lwd = 3)
> for(i in 1:5){
    fhat <- fk_sum(x, y, hs[i], x_eval = xeval) / fk_sum(x, rep(1, n),
        hs[i], x_eval = xeval)
    lines(xeval, fhat, lty = i, lwd = 2)
 }
\end{verbatim}
\begin{figure}
    \centering
    \includegraphics[height=5cm, width = 10cm]{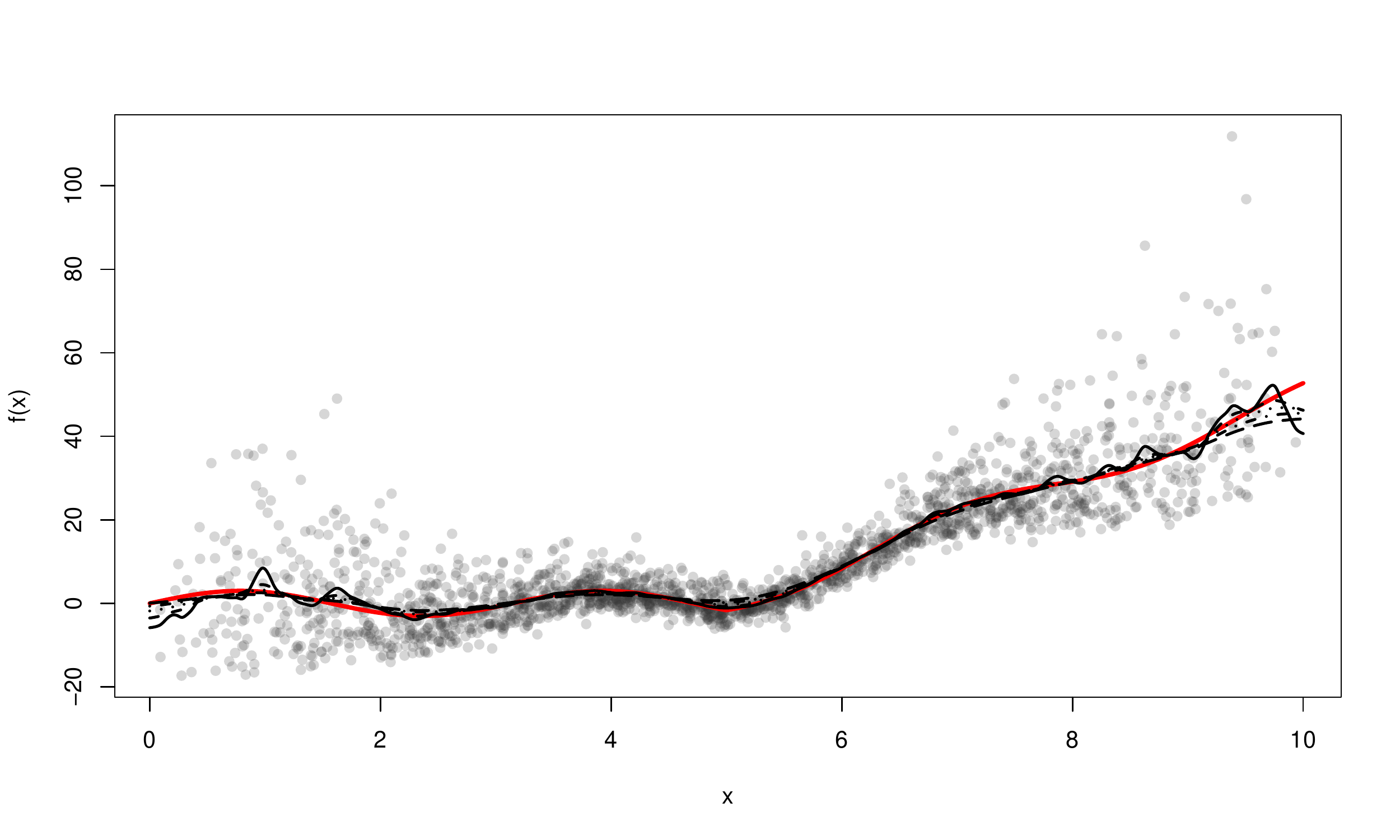}
    \includegraphics[height=5cm, width = 5cm]{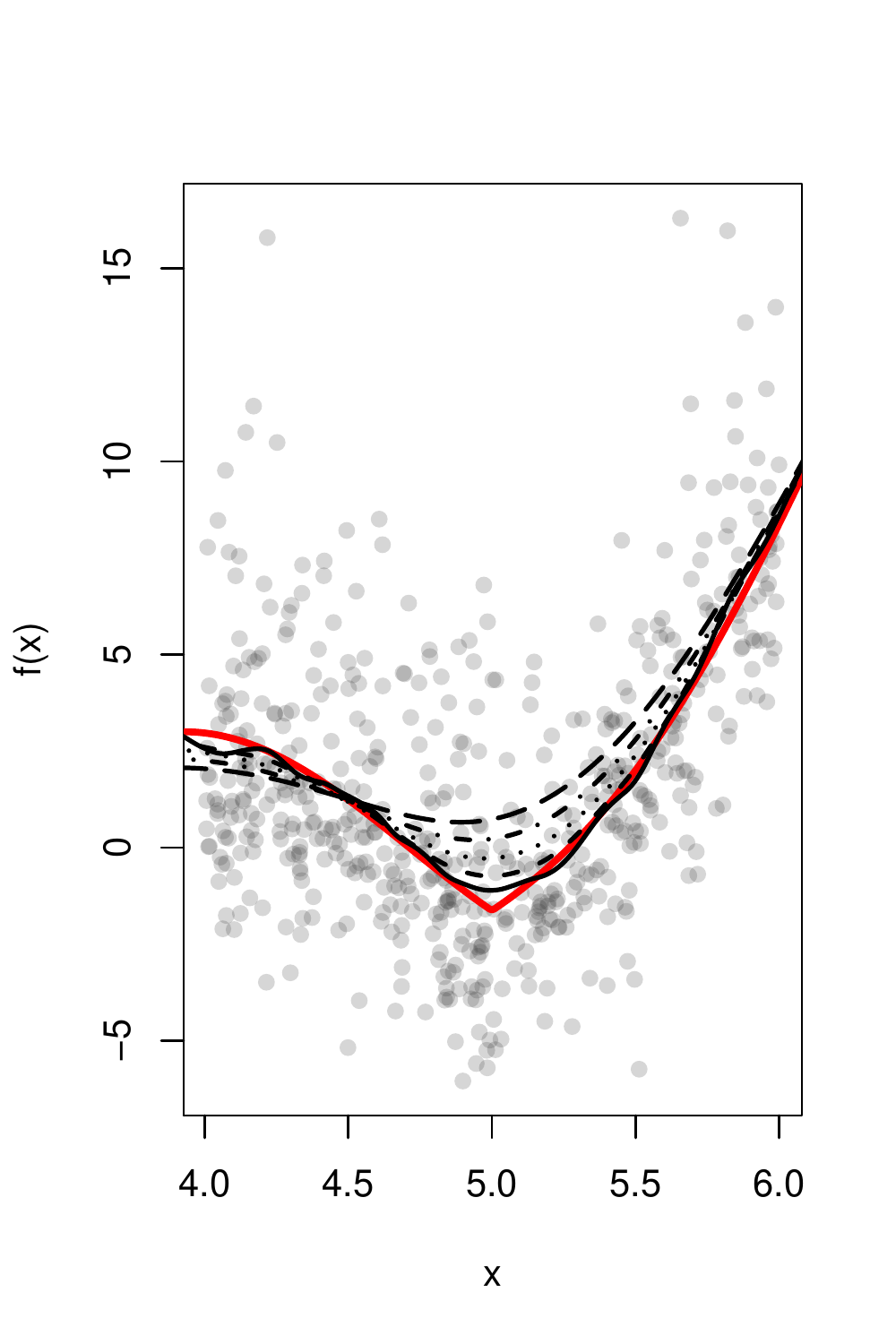}
    \caption{Nadaraya-Watson regression estimates of non-linear function (left) and zoomed middle section of function (right). True function shown in red, with estimates using various bandwidths shown with ------ ($h = 0.05$), - - - - ($h = 0.1$), $\cdots\cdots$ ($h = 0.15$), -$\cdot$-$\cdot$-$\cdot$- ($h = 0.2$), --~--~--~($h = 0.25$)}
    \label{fig:NWreg1}
\end{figure}
The results are shown in Figure~\ref{fig:NWreg1}. The true function is shown in red, and the estimates in black. Again the different estimates are similar over much of the range of the function. In this case, however, the variation of the smaller bandwidth estimates is very apparent near the limits of the range. This is due to the fact that the smaller bandwidth estimates have a lower effective sample size in the point-wise estimates of the function, and hence high variation where the data are sparse. A limitation of this particular estimator is also seen in the right extreme of the function, where the true function is underestimated. This is due to the bias of the Nadaraya-Watson (NW) estimator at the limits of the data if the gradient is non-zero. A similar bias is also seen in local extrema of the function, such as that shown in the right plot, which contains a zoomed view of the middle section of the function. Local-linear kernel regression has considerably lower bias in these extrema. The \pkg{FKSUM} package offers both NW and local-linear regression estimation with the function \code{fk\_regression()}. Details can be obtained with the command \code{help(fk\_regression)}. We chose to illustrate the NW estimator due to its simplicity, and as it provides a simple and intuitive application of the \code{fk\_sum()} function.

As before we can use cross-validation to select an appropriate bandwidth value based on the data. Arguably the most justifiable loss function with respect to which to perform cross validation in the context of regression is the squared loss. This is because the function which minimises $E_{X,Y}[(Y-g(X))^2]$ is the assumed regression function $g(x) = f(x) = E[Y|X=x]$. We therefore use the following loss function to perform cross validation.
%
%
%
%
\begin{verbatim}
> loo_sse <- function(h, x, y, beta){
    n <- length(x)
    numerator <- fk_sum(x, y, h, beta = beta) - beta[1] * y
    denominator <- fk_sum(x, rep(1, n), h, beta = beta) - beta[1]
    yhat <- numerator / denominator
    sum((y - yhat)^2)
 }
\end{verbatim}
That is, we compute the leave-one-out estimates of the numerator and denominator terms in the estimated function values, \code{yhat}. We then simply compute the sum of the squared deviations of these estimates from the observed responses.
We can then minimise the cross validation error, and use the corresponding bandwidth in estimating the regression function.

\begin{verbatim}
> h_cv <- optimise(loo_sse, c(.05, .5), x, y, c(.25, .25))$minimum
> h_cv


[1] 0.1152742


> fhat_cv <- fk_sum(x, y, h_cv, x_eval = xeval) / fk_sum(x, rep(1, n),
    h_cv, x_eval = xeval)
>
> plot(x, y, col = rgb(.2, .2, .2, .2), pch = 16, xlab = "x", ylab = "f(x)")
> lines(xeval, 3 * sin(2 * xeval) + 10 * (xeval > 5) * (xeval - 5),
      col = 2, lwd = 3)
> lines(xeval, fhat_cv, lwd = 2)
\end{verbatim}

\begin{figure}
    \centering
    \includegraphics[height=5cm, width = 10cm]{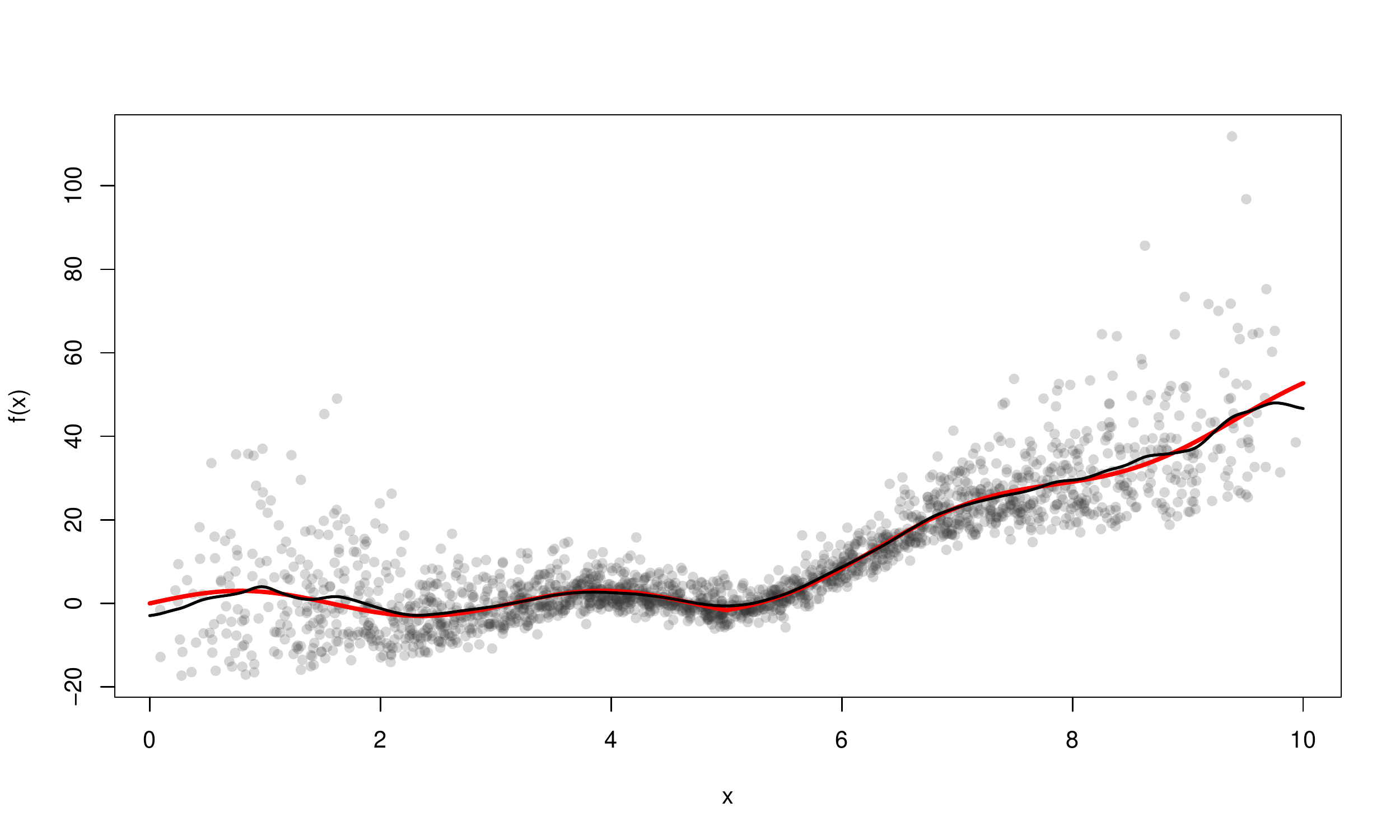}
    \includegraphics[height=5cm, width = 5cm]{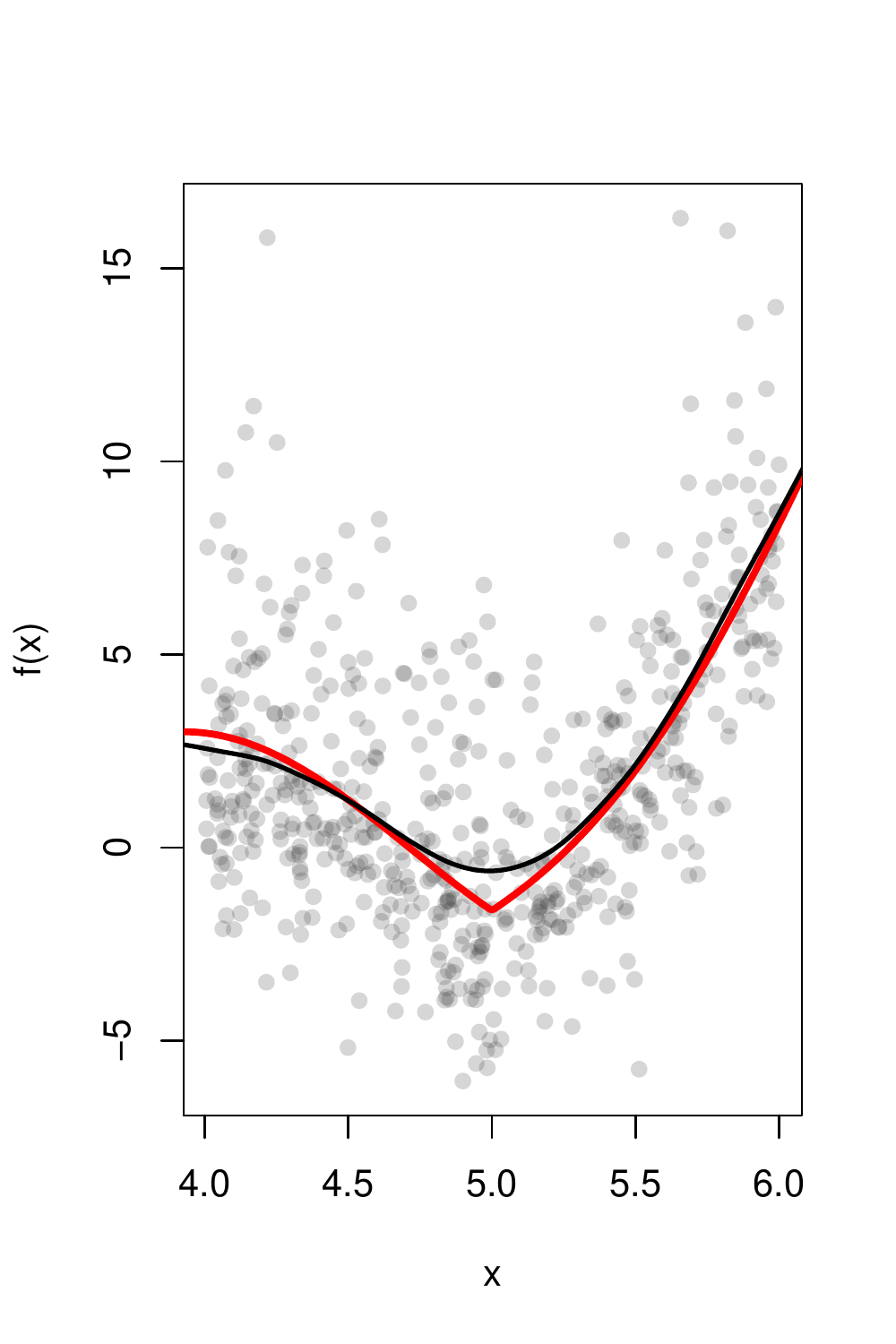}
    \caption{Nadaraya-Watson regression estimates of non-linear function (left) and zoomed middle section of function (right). True function shown in red, with estimate using bandwidth selected via cross-validation shown with ------}
    \label{fig:NWreg1}
\end{figure}
The selected bandwidth is close to the middle value used before, and captures the sharp local minimum in the middle of the range reasonably well, without being overly variable in the tails.\\
\\
In this section we introduced the versatile function \code{fk\_sum()}, which can be used to implement standard kernel based estimators very simply. Although the examples considered, being kernel density and regression function estimation, are also implemented in the purpose-built functions \code{fk\_density()} and \code{fk\_regression()}, we hope that these examples have given the reader sufficient understanding of the function's use that they will be capable of implementing their own simple kernel estimators. In the next section we discuss some of the functionality offered by the package for projection pursuit problems. In addition to two purpose-built implementations, we discuss in detail the implementation of projection pursuit regression using the Nadaraya-Watson estimator. This example should give more advanced readers further instruction on the use of the function \code{fk\_sum()} in more complex problems.


\section{Projection Pursuit}\label{sec:pp}

Projection pursuit refers to the problem of identifying (usually low-dimensional) linear projections of a set of data which expose structures of interest. What is considered interesting may be made explicit in relation to a subsequent objective, e.g., clustering~\citep{bolton2003projection,hofmeyr2019RJ}, or the pursuit may be more exploratory in nature~\citep{huber1985projection,friedman1987exploratory}. Projection pursuit may also be formulated for supervised problems, such as regression~\citep{friedman1981projection}, in which projections of a collection of covariates are sought which reveal strong predictive relationships with a set of response variables. Arguably the most popular projection pursuit method is that of Principal Component Analysis (PCA). PCA has numerous equivalent formulations, including finding the projection of the data which minimises the total squared reconstruction error. In the context of projection pursuit this may be seen rather as the objective of finding projections which lose as little structure as possible, in a general sense. However, PCA may fail in this objective if the variation in the noisy (less structured) components in the data dominates the total data variation. 
%

The basic formulation of the projection pursuit problem may be given as

\begin{align*}
    \max_{\W \in \mathcal{F}} \Phi(\W | \X),
\end{align*}
where $\X \in \R^{n\times d}$ represents the data matrix, $\W$ is the {\em projection matrix}, which comes from some feasible set $\mathcal{F} \subset \R^{d\times d'}$, and $\Phi(\cdot)$ is the {\em projection index}, which is intended to measure how ``interesting'' are the projected data, $\X\W$. That is, we seek the projection matrix $\W$ on which the projected data, $\X\W\in \R^{n\times d'}$, capture the structures of interest (which are in $\X$) as well as possible. 
%
%
Very frequently an equivalent formulation of the projection index allows us to decompose the total ``interestingness'' of the projected data into the sum of the interestingness of each of the {\em components}, $\X\w_i, i = 1, ..., d'$, measured separately. Even if the reformulation is not equivalent, this latter approach is frequently preferred as it can vastly simplify the problem. The main practical benefit of this reformulation is that it allows us to obtain the {\em projection vectors} (i.e., the columns of $\W$) iteratively. The feasible set $\mathcal{F}$ is then used (either explicitly or implicitly) to effectively ensure that the same, or very similar vectors are not obtained for multiple columns of $\W$.
%
%
With this reformulation we consider the projection pursuit problem expressed, somewhat loosely, as,
\begin{align*}
    \max_{\W : \w_i \in \mathcal{F}(\W_{-i})}\sum_{i=1}^{d'}\Phi\left(\w_i | \X, \W_{-i}\right),
\end{align*}
where $\W_{-i}$ denotes the matrix $[\w_1 \ ... \ \w_{i-1} \ \ \w_{i+1} \ ... \ \w_{d'}]$, i.e., the projection matrix excluding the $i$-th column.
That is, we make explicit the fact that the feasible set, and indeed the projection index itself, for the $i$-th component, may depend on the values of the other components. As mentioned above, the problem in this form is often approached in an iterative manner, in which the optimisation is greedily performed by adding the apparently best component at iteration $i$, given all those obtained so far, but without accounting for components which will be obtained subsequently. In this case we are essentially then interested only in the univariate projection pursuit problem,
\begin{align*}
    \max_{\w \in \mathcal{F}} \Phi(\w |\X),
\end{align*}
where we have suppressed the dependence on the other components for convenience of notation, and note that in order to obtain a complete solution we may have to consider a sequence of different problems, i.e., with different objectives and feasible sets. Now, except in examples in which the projection index is {\em scale invariant}, i.e., $\Phi(\w |\X) = \Phi(\alpha \w |\X) \ \forall \alpha > 0$, a fairly universal constraint is that the norm of the projection vector is constant, where without loss of generality we may assume $||\w|| = 1$. Enforcing this constraint can be achieved in a number of ways, including using modified search directions during optimisation~\citep{niu2011dimensionality}, expressing $\w$ in polar coordinates~\citep{HofmeyrP2015} or by formulating the projection index as the composition~\citep{Hofmeyr2016}
\begin{align*}
    \Phi(\w | \X) = \phi(\p)\bigg|_{\p = \X\vec \w},
\end{align*}
where $\vec\w = \frac{1}{||\w||}\w$ and $\phi(\cdot)$ simply measures the interestingness of a univariate data set, here given as the normalised projected data, stored in the vector $\p = (p_1, ..., p_n) = (\x_1^\top \vec \w, ..., \x_n^\top \vec \w)$. Practically, then, $\Phi(\cdot)$ preforms two operations. First it projects the argument $\w$ onto the unit ball (i.e., divides it by its norm), and then computes the interestingness of the data projected onto the unit vector $\vec\w$. Optimisation of $\Phi(\cdot)$ based on this formulation can thus equivalently be performed without any constraint on the magnitude of the argument $\w$.

There are three projection pursuit methods which are implemented in \mypkg. These are Independent Component Analysis~\citep[ICA]{HyvarinenO2000}, Minimum Density Projection Pursuit~\citep[MDPP]{pavlidis2016minimum} and Projection Pursuit Regression~\citep[PPR]{friedman1981projection}. We will discuss the purpose-built implementations of ICA and MDPP in this section. In addition, we provide a detailed discussion of the implementation of projection pursuit regression with the intention that this will provide instruction for readers who may wish to implement their own projection pursuit methods, or existing methods which were not implemented in the package.


\subsection{Independent Component Analysis}

A primary motivation for the application of independent component analysis is the assumption that an observed multivariate data set, $\X \in \R^d$, has as columns different linear combinations of latent factors which are statistically independent. The task is then to obtain an {\em unmixing matrix}, $\W$, such that the columns of the transformed data $\X\W$ have maximal independence. This has broad application in signal processing; with the canonical example being the so-called {\em cocktail party problem}. Here a room is populated by a crowd of people, with different subsets involved in different conversations. Microphones situated around the room record these conversations, with each microphone recording conversations closer to them with higher volume, but also receiving sounds from all other conversations in the room at different levels. The objective is then to take all recordings and identify a transformation which separates the actual conversations from one another.


A popular objective for ICA uncovers an alternative point of view which motivates the application of ICA to the general problem of exploratory projection pursuit. That is, to maximise independence one can minimise the mutual information in the components of the projected data. In order to achieve this, we attempt to maximise the unique information in each of the components, and so in a general sense ICA may be seen as obtaining the projection which identifies the most informative components in the data. Now, if we consider the statistical context, in which our observations represent a sample of realisations of a random variable, say $\Z$, which may without loss of generality be assumed to have zero mean, then the objective of minimising mutual information is to minimise the Kullback-Leibler divergence~\citep[KL-divergence]{KLdivergence} between the joint distribution of $\Z^\top\W$ and the product of the marginal distributions of $\Z^\top\w_i, i = 1, 2, ..., d'$.
That is, if we use $f_{\Z^\top\W}(\cdot)$ and $f_{\Z^\top \w_i}(\cdot)$ to represent the densities of $\Z^\top \W$ and $\Z^\top \w_i$ respectively, then
\begin{align*}
    KL\left(f_{\Z^\top\W} \bigg|\bigg| \prod_{i=1}^{d'} f_{\Z^\top\w_i}\right) &= E\left[\log(f_{\Z^\top \W}(\Z^\top \W))\right] - E\left[\log\left(\prod_{i=1}^{d'} f_{Z^\top \w_i}(\Z^\top \w_i)\right)\right]\\
    &= E[f(\Z)] + \log(\mbox{det}|\W|) - \sum_{i=1}^{d'} E[\log(f_{\Z^\top\w_i}(\Z^\top \w_i))].
\end{align*}
If this divergence is small, then the joint distribution is well approximated by the product of its marginals, meaning that the corresponding random variables are close to independent. Now, two useful observations vastly simplify the problem of estimating an appropriate projection matrix, $\W$, to achieve this minimisation. First, a necessary condition for two random variables to be independent is that they have zero correlation. Second, independence is unaffected by scale, in that two random variables, say $U, V$ are independent if and only if $U$ and $\alpha V$ are independent for all $\alpha$. Practically, then, we restrict $\W$ to be of the form $\pmb{\Lambda}^{-1/2}\mathbf{U}\mathbf{Q}$, where $\pmb{\Lambda}$ is the diagonal matrix containing the eigenvalues of the covariance of $\Z$, $\mathbf{U}$ has as columns its first $d'$ eigenvectors, and $\mathbf{Q} \in \R^{d'\times d'}$ is orthonormal. The convenience of this is that the determinant of $\W$ is constant, meaning that the dependence of the above KL-divergence on $\W$ is only in the term $\sum_{i=1}^{d'}E[\log(f_{\Z^\top\w_i}(\Z^\top\w_i))]$. This is the sum of the negative entropies of the random variables $\Z^\top \w_1, ..., \Z^\top \w_{d'}$. Notice also that the term $\pmb{\Lambda}^{-1/2}\mathbf{U}$ has the effect of whitening the random variable, so that $\Z\pmb{\Lambda}^{-1/2}\mathbf{U}$ has identity covariance. Practically, then, to perform ICA, we first whiten the data matrix by setting $\tilde \X = (\X - \hat{\pmb{\mu}}\one^\top)\hat{\pmb{\Lambda}}^{-1/2}\hat{\mathbf{U}}$, where $\hat{\pmb{\Lambda}}$ and $\hat{\mathbf{U}}$ contain the eigenvalues and vectors of the covariance matrix of the data, and $\hat{\pmb{\mu}}$ is the vector of column means of the data. We then iteratively minimise the estimated sample entropy of the projected components, with the $i$-th minimisation being of
\begin{align*}
    -\frac{1}{n}\sum_{j=1}^n \log\left(\hat f_{\p}(p_j)\right)\bigg|_{\p = \tilde\X\vec{\mathbf{q}_i}},
\end{align*}
over the projection vector, now denoted by $\mathbf{q}_i$, under the constraint that $\mathbf{q}_i$ is orthogonal to $\mathbf{q}_1, ..., \mathbf{q}_{i-1}$. Here $\mathbf{q}_i$ is, after normalisation, the $i$-th column of the orthonormal matrix $\mathbf{Q}$. We use $\hat f_{\p}(\cdot)$ to be the estimated density based on the vector of univariate projected points, where in this case this is given by the  projection of the whitened data, $\tilde\X$, onto the normalised projection vector $\vec{\mathbf{q}_i}$. The natural choice of density estimate in our context is the kernel density estimate, discussed previously.

The implementation of ICA in \mypkg~is provided in the function \code{fk\_ICA()}, and is based on the above approach. The function takes the following arguments:
\begin{align*}
    \mbox{\code{X}}:& \mbox{ matrix of observations (num\_data x num\_variables) } \X.\\
    \mbox{\code{ncomp}}:& \mbox{ (optional) integer number of independent components to extract. The default is 1.}\\
    \mbox{\code{beta}}:& \mbox{ (optional) vector of kernel coefficients. The default is \code{beta = c(0.25, 0.25)}.}\\
    \mbox{\code{hmult}}:& \mbox{ (optional) bandwidth multiplier. Bandwidth used is Silverman's rule}\\
    & \mbox{ multiplied by \code{hmult}. The default is 1.5.}\\
    \mbox{\code{it}}:& \mbox{ (optional) integer number of iterations in each optimisation. The default is 20.}\\
    \mbox{\code{nbin}}: & \mbox{ (opitonal) integer number of bins if binning approximation is to be used.}\\
    & \mbox{ The default is to perform exact kernel computations.}
\end{align*}

The output of the function is a list with the following fields:
\begin{align*}
    \mbox{\code{\$X}}:& \mbox{ the data matrix given as argument to the function.}\\
    \mbox{\code{\$K}}:& \mbox{ the whitening matrix, } \hat{\pmb{\Lambda}}^{-1/2}\hat{\mathbf{U}}.\\
    \mbox{\code{\$W}}:& \mbox{ the optimal projection matrix for the whitened data.}\\
    \mbox{\code{\$S}}:& \mbox{ The estimated independent components.}
\end{align*}

Note that in the package we denote the unmixing matrix for the already whitened data by \code{\$W} to be consistent with other implementations, where in our derivation above we used $\mathbf{Q}$ to represent this matrix.

\paragraph{Example: Simulation}

In the first example using the \code{fk\_ICA()} function we conduct a simulation study to assess the speed and accuracy of the implementation in recovering components which are simulated under the independence assumption, but which are then subjected to a random linear transformation using a randomly generated mixing matrix. Following~\cite{KernelICA} and~\cite{ProDenICA}, we use the Amari distance~\citep{amari} between the estimated unmixing matrix and the inverse of the mixing matrix to assess the accuracy of the estimation. We also use the collection of densities introduced by~\cite{KernelICA} to simulate the true independent components.
For context, we compare with the ICA methods implemented in the packages~\pkg{fastICA}~\citep{CRANfastICA}, \pkg{ProDenICA}~\citep{CRANProDenICA} and \pkg{PearsonICA}~\citep{CRANPearsonICA}. We begin by installing these packages.
\begin{verbatim}
> if(! "fastICA" %in% installed.packages()) install.packages("fastICA")
> if(! "ProDenICA" %in% installed.packages()) install.packages("ProDenICA")
> if(! "PearsonICA" %in% installed.packages()) install.packages("PearsonICA")
>
> library(fastICA)
> library(ProDenICA)
> library(PearsonICA)
\end{verbatim}
The \pkg{ProDenICA} package also conveniently provides functions for simulating data from the densities in~\cite{KernelICA} and for computing the Amari distance, in the functions \code{rjordan()} and \code{amari()} respectively. It also provides the function \code{mixmat()} for simulating mixing matrices.

We compare the methods based on the accuracy of their outputs, and in terms of their running time. We repeatedly generate sets of data with independent components, and then apply a randomly generated linear transformation to these data and apply the various ICA methods. We repeat the experiment 50 times, and in each case sample 2000 points with 4 components.
\begin{verbatim}
> n_rep <- 50
> n_dat <- 2000
> n_components <- 4
\end{verbatim}
We then store the quality of the solutions, and the running times, for each of the methods in the following vectors.
\begin{verbatim}
> amari_fk <- numeric(n_rep)
> amari_fast <- numeric(n_rep)
> amari_Pearson <- numeric(n_rep)
> amari_ProDen <- numeric(n_rep)
>
> t_fk <- numeric(n_rep)
> t_fast <- numeric(n_rep)
> t_Pearson <- numeric(n_rep)
> t_ProDen <- numeric(n_rep)
\end{verbatim}
Within each experiment we choose a random combination of the 18 densities produced by the \code{rjordan()} function, which are listed according to the letters \code{`a'} to \code{`r'}.
%
%
In the following, in each experiment the matrix \code{X} contains the true components and \code{R} the random mixing matrix. \code{Xx} therefore contains the mixed components. We then apply each of the methods and store the resulting performance measures.
\begin{verbatim}
> for(rep in 1:n_rep){
      set.seed(rep)
      
      X <- matrix(0, n_dat, n_components)
      densities <- sample(letters[1:18], n_components)
      for(i in 1:n_components)  X[,i] <- rjordan(densities[i], n_dat)
      R <- mixmat(n_components)
      Xx <- X %*% R
      
      t_fk[rep] <- system.time(model <- fk_ICA(Xx, n_components))[1]
      amari_fk[rep] <- amari(model$K %*% model$W, solve(R))
      
      t_fast[rep] <- system.time(model <- fastICA(Xx, n_components))[1]
      amari_fast[rep] <- amari(model$K %*% model$W, solve(R))
      
      t_ProDen[rep] <- system.time(model <- ProDenICA(Xx, n_components,
        W0 = diag(n_components), whiten = TRUE))[1]
      amari_ProDen[rep] <- amari(model$whitener %*% model$W, solve(R))
      
      t_Pearson[rep] <- system.time(model <- PearsonICA(Xx, n_components,
        w.init = diag(n_components)))[1]
      amari_Pearson[rep] <- amari(model$W, solve(R))
 }
\end{verbatim}
In the above we initialised both \code{ProDenICA()} and \code{PearsonICA()} with the identity matrix. The reason is that these functions initialise with a random matrix, where the random number generation is not conducted within {R}, and hence does not respect the random number seed setting in {R}. This initialisation is the same as that used in the \code{fk\_ICA()} function.

Next we compare the mean Amari distance and running time of the methods.
\begin{verbatim}
> colMeans(cbind(amari_fk, amari_fast, amari_ProDen, amari_Pearson))
     amari_fk    amari_fast  amari_ProDen amari_Pearson 
   0.08467769    0.15133353    0.04919577    0.18757256
> colMeans(cbind(t_fk, t_fast, t_ProDen, t_Pearson))
     t_fk    t_fast  t_ProDen t_Pearson 
  0.14328   0.01228   2.17880   0.03022 
\end{verbatim}
The implementation in \pkg{ProDenICA} obtains the most accurate results, but runs in considerably longer time than the other methods. The running time will clearly depend on the system being used, but we expect a similar relative running time for comparison on all systems. Both \pkg{fastICA} and \pkg{PearsonICA} are very computationally efficient, but they are much less reliable in accurately recovering the underlying components. The method in \mypkg~offers a reasonable trade-off; achieving close-to as accurate outputs to those of \pkg{ProDenICA}, but far more efficiently, on these data.

 It is worth noting that the method in \pkg{ProDenICA} uses a binning approximation to estimate the densities of the components during optimisation. As a result, we expect this method will compare less unfavourably in terms of running time for larger data sets. In the next example we consider such a larger problem. We also add the binning based approximate version of \code{fk\_ICA()} for comparison.

\paragraph{Example: The Cocktail Party Problem} 

Here we follow Example 1 of~\cite{CRANJADE}, but repeat it multiple times. The example uses three audio waves of length 50000 and a single noise signal as the original components, and then applies a random mixing matrix to obtain the observed mixed signals. We begin by loading the packages required for this example, and then loading the audio data.
\begin{verbatim}
> if(! "JADE" %in% installed.packages()) install.packages("JADE")
> library(JADE)
> if(! "tuneR" %in% installed.packages()) install.packages("tuneR")
> library(tuneR)
> S1 <- readWave(system.file("datafiles/source5.wav", package = "JADE"))
> S2 <- readWave(system.file("datafiles/source7.wav", package = "JADE"))
> S3 <- readWave(system.file("datafiles/source9.wav", package = "JADE"))
\end{verbatim}
Next we repeatedly generate a random noise signal and mixing matrix, and fit the same models as in the last example,  with the addition of the binning version of \code{fk\_ICA()}. We compute the amari distance between the estimated unmixing matrices and the inverse of the mixing matrix, \code{R}. We also store the running times of the different methods. Because the running times are considerably greater on these larger data, we only repeat the experiment 20 times. 
%
%
\begin{verbatim}
> n_rep <- 20
> 
> amari_fk <- numeric(n_rep)
> amari_fk_bin <- numeric(n_rep)
> amari_fast <- numeric(n_rep)
> amari_Pearson <- numeric(n_rep)
> amari_ProDen <- numeric(n_rep)
>
> t_fk <- numeric(n_rep)
> t_fk_bin <- numeric(n_rep)
> t_fast <- numeric(n_rep)
> t_Pearson <- numeric(n_rep)
> t_ProDen <- numeric(n_rep)
> 
> for(rep in 1:n_rep){
    set.seed(rep)
    NOISE <- noise("white", duration = 50000)
    X <- cbind(S1@left, S2@left, S3@left, NOISE@left)
    R <- mixmat(4)
    Xx <- X %*% R

    t_fk[rep] <- system.time(model <- fk_ICA(Xx, 4))[1]
    amari_fk[rep] <- amari(model$K %*% model$W, solve(R))
      
    t_fk_bin[rep] <- system.time(model <- fk_ICA(Xx, 4, nbin = 5000))[1]
    amari_fk_bin[rep] <- amari(model$K %*% model$W, solve(R))
      
    t_fast[rep] <- system.time(model <- fastICA(Xx, 4))[1]
    amari_fast[rep] <- amari(model$K %*% model$W, solve(R))
      
    t_ProDen[rep] <- system.time(model <- ProDenICA(Xx, 4, W0 = diag(4),
        whiten = TRUE))[1]
    amari_ProDen[rep] <- amari(model$whitener %*% model$W, solve(R))
      
    t_Pearson[rep] <- system.time(model <- PearsonICA(Xx, 
        w.init = diag(4), 4))[1]
    amari_Pearson[rep] <- amari(model$W, solve(R))
 }
> colMeans(cbind(amari_fk, amari_fk_bin, amari_fast, amari_ProDen,
    amari_Pearson))
     amari_fk  amari_fk_bin    amari_fast  amari_ProDen amari_Pearson 
    0.1416785     0.1392952     0.1605889     0.1854821     0.1866236  
> colMeans(cbind(t_fk, t_fk_bin, t_fast, t_ProDen, t_Pearson))
     t_fk  t_fk_bin    t_fast  t_ProDen t_Pearson 
  2.43345   0.38030   0.07330   4.98875   0.21880 
\end{verbatim}
%
%
Again, the running times may differ substantially based on the system being used, but the relative running times should be similar regardless. On this larger data set, the method in \pkg{ProDenICA} is only roughly twice as slow as \code{fk\_ICA()}. When the binning approximation is used in combination with the fast kernel methods in \mypkg, however, \code{fk\_ICA()} runs roughly ten times faster than \code{ProDenICA()}. Again \code{fastICA()} and \code{PearsonICA()} are very computationally efficient.

In this example \code{fk\_ICA()} obtains the most accurate results. In fact the binning approximation yields a very slightly more accurate output than the exact kernel method. Here the performance of \code{fastICA()} is only slightly worse than \code{fk\_ICA()}, while \code{PearsonICA()} and \code{ProDenICA()} are the least accurate.

\subsection{Minimum Density Hyperplanes}

The minimum density projection pursuit method~\citep{pavlidis2016minimum} attempts to obtain the cluster separating hyperplane with minimal integrated density along it; the Minimum Density Hyperplane (MDH). That is, to obtain the hyperplane $H(\w, b):= \{\x \in \R^d | \w^\top\x = b\}$ for which the surface integral on the hyperplane, $\oint_{H(\w,b)} f(\x)d\x$, is minimised. Here $f(\cdot)$ is a probability density function, and this integral is, in practice, estimated using a kernel estimate based on a sample from the distribution with density $f(\cdot)$. Notice that if $X$ is a random variable with density $f(\cdot)$, then $\oint_{H(\w,b)} f(\x)d\x$ is equal to the value of the density of the univariate random variable $X^\top\vec\w$ evaluated at $\frac{b}{||\w||}$. To estimate this integral, therefore, one needs only compute a univariate density estimate from the projected data on $\vec \w$. Now, it should be clear that the integral on the hyperplane $H(\w,b)$ can be made arbitrarily small by taking a value of $b$ which is extremely large in magnitude. To ensure the hyperplane passes through the region of the density which is of interest, rather than through the tail, a penalty is added to the objective which ensures the hyperplane passes within a chosen distance of the mean. In particular, the objective is given by
\begin{align}\label{eq:mdh_obj}
    p(\w, b) = \widehat{\oint_{H(\w,b)} f(\x)d\x} + Cd\left(b, [\hat\mu_{\w}-\alpha \hat\sigma_{\w},\hat \mu_{\w}+\alpha \hat\sigma_{\w}]\right)^2,
\end{align}
where $d(\cdot,\cdot)$ is the Euclidean metric, and $\hat\mu_{\w} = \w^\top \hat{\pmb{\mu}}$ and $\hat \sigma_{\w} = \sqrt{\w^\top \hat\Sigma \w}$ are the mean and standard deviation of the projected data, $\X\w$, respectively. The constant $C > 0$ is some large positive value which affects the strength of the penalty and $\alpha > 0$ controls the size of the feasible region. The second term in Eq.~(\ref{eq:mdh_obj}) therefore applies no penalty for hyperplanes which pass within $\alpha$ standard deviations of the mean of the data, measured in the the direction orthogonal to the hyperplane. Beyond the distance $\alpha$, the penalty scales quadratically with the distance of the hyperplane from the mean. Especially when the number of clusters is large, the number of local minima in the objective in Eq.~(\ref{eq:mdh_obj}) tends to be large. To mitigate the effect of these minima, the Minimum Density Projection Pursuit (MDPP) objective is given by,
\begin{align}\label{eq:MDPP}
    \Phi(\w | \X) := \min_{b\in \R} p(\vec \w, b).
\end{align}
Practically, for a given $\w$, we compute a kernel density estimate from the projected data on the normalised projection vector, $\X\vec\w$, and search for the minimum of this density plus the penalty described in Eq.~(\ref{eq:mdh_obj}). In other words, the projection index for projection vector $\w$ is the value of the original objective for the best hyperplane orthogonal to $\w$. This approach allows the value of $b$ to change discontinuously during optimisation, and thereby avoid some of the local minima in the original objective. The hyper-parameter $\alpha$ is adjusted during optimisation. Starting from $\alpha = 0$, the first solution passes through the mean of the data, and tends to lead to a reasonable separation of clusters. Thereafter the value is slowly increased and the solution returned by the algorithm is the last {\em valid} cluster separator (a hyperplane which passes between the modes of the density of the projected data).

MDPP is implemented in the function \code{fk\_mdh()}, which takes the following arguments:
\begin{align*}
    \mbox{\code{X}}:& \mbox{ matrix of observations (num\_data $\times$ num\_variables) } \X.\\
    \mbox{\code{v0}}:& \mbox{ (optional) initial projection vector. The default is the first principal component.}\\
    \mbox{\code{hmult}}:& \mbox{ (optional) numeric bandwidth multiplier. The bandwidth used to estimate the}\\
    & \mbox{ density is given by \code{hmult} multiplied by Silverman's heuristic calculated on the}\\
    & \mbox{ initial projection vector. The default value is 1.}\\
    \mbox{\code{beta}}:& \mbox{ (optional) vector of kernel coefficients. The default is \code{beta = c(0.25, 0.25)},}\\
    & \mbox{ corresponding to the smooth order 1 kernel.}\\
    \mbox{\code{alphamax}}:& \mbox{ (optional) numeric maximum (scaled) distance of the hyperplane from the mean}\\
    & \mbox{ of the data. The default is 1.}
\end{align*}
The function returns a list with fields \code{\$v} and \code{\$b}, corresponding to the final (normalised) projection vector, $\vec\w$, and the value of $b$ in the optimal hyperplane orthogonal to $\vec \w$.

It is worth noting that in order to compute $\Phi(\w|\X)$, the search for the optimal hyperplane orthogonal to $\w$ is performed using a combination of grid evaluations and binary search. As a result, there is no need to evaluate the density at every projected data point, as was the case in ICA. We therefore do not expect so significant an improvement on the running time over existing implementations. That being said, repeated evaluation of the density during binary search is more efficient using the fast kernel computations availed by the package.

\paragraph{Example: Simulation}
We begin, as before, with a simulation to assess the speed and accuracy of the implementation in \mypkg. We compare with the implementation given in \pkg{PPCI}~\citep{hofmeyr2019RJ}, which uses the Gaussian kernel for estimating densities. We begin by loading the corresponding library.
\begin{verbatim}
> if(! "PPCI" %in% installed.packages()) install.packages("PPCI")
> library(PPCI)
\end{verbatim}
In each experiment we simulate 2000 data from a 10 component Gaussian mixture in 10 dimensions, with means and diagonal covariances determined randomly. We will measure the accuracy in terms of the true integrated density on the hyperplane solutions, as well as the clustering accuracy of the separation of the collections of points generated from each of the mixture components. Since a hyperplane only induces a binary partition of the data, we evaluate the clustering accuracy using the success ratio~\citep{pavlidis2016minimum}, which measures the degree to which at least one cluster has been successfully separated from the remainder. We repeat the experiment 100 times, and store the resulting running times and accuracy measures.
\begin{verbatim}
> n_dat <- 2000
> n_dim <- 10
> n_comp <- 10
> n_rep <- 100
>
> t_fk <- numeric(n_rep)
> t_pp <- numeric(n_rep)
> dens_int_fk <- numeric(n_rep)
> dens_int_pp <- numeric(n_rep)
> s_ratio_fk <- numeric(n_rep)
> s_ratio_pp <- numeric(n_rep)
\end{verbatim}
To simulate data from a Gaussian mixture, we begin by generating a matrix of means (\code{mu}) whose rows represent the means of the mixture components, and a matrix of standard deviations (\code{sds}) whose rows are the square roots of the diagonal elements of the covariance matrices of the components. We also simulate a vector of mixing proportions (\code{ps}).
To generate data from this mixture model, we simulate the cluster indicator matrix (\code{I}) which is the binary matrix with a one in position $i,j$ if an only if point $i$ is from cluster (component) $j$. We then compute the matrix of point means, \code{I \%*\% mu}, and the matrix of residuals, whose $i,j$-th element is of the form \code{rnorm(1) * I[i, ] \%*\% sds[ ,j]}. We apply both \code{fk\_mdh()} and the function \code{mdh()} from the package \pkg{PPCI} to the resulting data matrix, \code{X}, and compute the performance measures.
\begin{verbatim}
> for(rep in 1:n_rep){
    set.seed(rep)
    
    mu <- matrix(runif(n_dim * n_comp), n_comp, n_dim)
    sds <- matrix(rexp(n_dim * n_comp), n_comp, n_dim) / 7
    ps <- runif(n_comp) + .1
    ps <- ps / sum(ps)

    I <- t(rmultinom(n_dat, 1, ps))
    
    M <- I %*% mu
    R <- matrix(rnorm(n_dat * n_dim), n_dat, n_dim) * (I %*% sds)
    X <- M + R

    t_fk[rep] <- system.time(model <- fk_mdh(X))[1]
    dens_int_fk[rep] <- sum(ps * dnorm(model$b, mu %*% model$v, 
        model$v %*% (sds * model$v)))
    s_ratio_fk[rep] <- success_ratio(X %*% model$v < model$b[1],
        apply(I, 1, which.max))

    t_pp[rep] <- system.time(model <- mdh(X))[1]
    dens_int_pp[rep] <- sum(ps * dnorm(model$b, mu %*% model$v,
        model$v %*% (sds * model$v)))
    s_ratio_pp[rep] <- success_ratio(X %*% model$v < model$b[1],
        apply(I, 1, which.max))
 }
\end{verbatim}
Finally we compare the average performance.
\begin{verbatim}
> colMeans(cbind(t_fk, t_pp))
   t_fk    t_pp 
0.12317 0.51878 
> colMeans(cbind(dens_int_fk, dens_int_pp))
dens_int_fk dens_int_pp 
  0.1996598   0.1941097
> colMeans(cbind(s_ratio_fk, s_ratio_pp))
s_ratio_fk s_ratio_pp 
 0.9210795  0.9286993
\end{verbatim}
Based on these measures, we see a very minor decrease in performance compared with the existing implementation, but a substantial computational gain, decreasing the average running time by roughly four times, on average.

\begin{figure}[h]
    \centering
    \includegraphics[width = 11cm, height = 8cm]{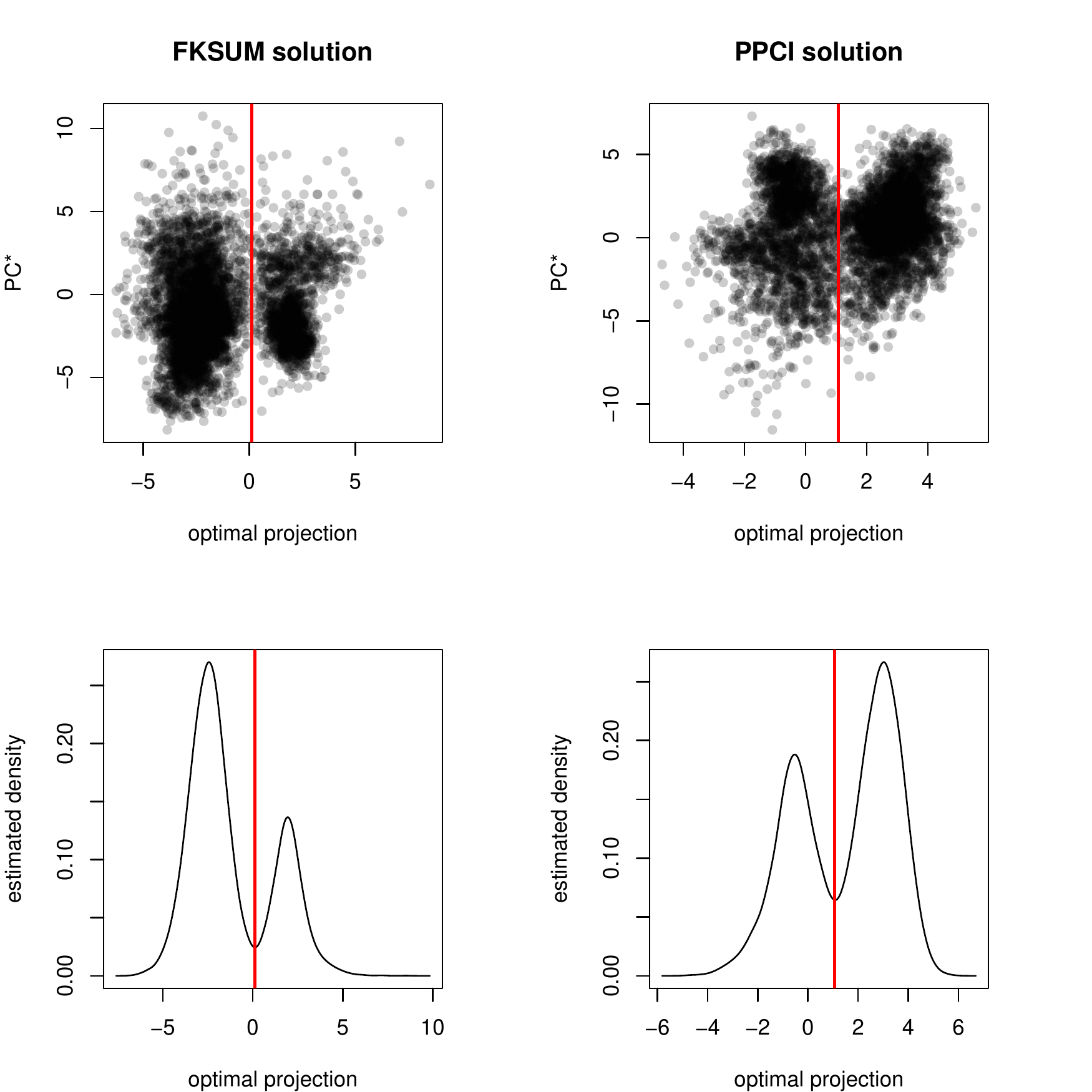}
    \caption{Minimum density hyperplane solutions on Optical Recognition of Handwritten Digits data. The top plots show the data projected into a two-dimensional subspace determined by the optimal projection vector and the first principal component in its null-space, while the bottom plots show the density of the data projected on the optimal projection vector.}
    \label{fig:optidigits_mdh}
\end{figure}

\paragraph{Example: Digit recognition} Next we apply both implementations of MDPP to two publicly available data sets, both of which are available in the \pkg{PPCI} package, and which were originally obtained from the UCI machine learning repository~\citep{uci}. The first is the ``Optical Recognition of Handwritten Digits'' data set, in which 5620 images of handwritten digits from multiple subjects have been compressed to 8$\times$8 pixels and vectorised, resulting in 64 dimensional data. We load this data set and apply both implementations.
\begin{verbatim}
> data("optidigits")
>
> system.time(fk_sol <- fk_mdh(optidigits$x))
   user  system elapsed 
  3.689   0.057   3.746
> success_ratio(optidigits$x %*% fk_sol$v < fk_sol$b[1], optidigits$c)
[1] 0.9299176
> system.time(pp_sol <- mdh(optidigits$x))
   user  system elapsed 
  5.318   0.157   5.476
> success_ratio(optidigits$x %*% pp_sol$v < pp_sol$b[1], optidigits$c)
[1] 0.9040637
\end{verbatim}
In this case the method in \mypkg~offers only a minor improvement in running time, but obtains a superior clustering result. Next we plot the resulting hyperplanes, and the estimated density along the optimal projection vectors.
\begin{verbatim}
> par(mfrow = c(2, 2))
>
> v2 <- eigen(cov(optidigits$x - 
    optidigits$x %*% fk_sol$v %*% t(fk_sol$v)))$vectors[,1]
> plot(optidigits$x %*% fk_sol$v, optidigits$x %*% v2, col = rgb(0, 0, 0, .2), 
    xlab = "optimal projection", ylab = "PC*", main = "FKSUM solution",
    pch = 16)
> abline(v = fk_sol$b, col = 2, lwd = 2)
> v2 <- eigen(cov(optidigits$x - 
    optidigits$x %*% pp_sol$v %*% t(pp_sol$v)))$vectors[,1]
> plot(optidigits$x %*% pp_sol$v, optidigits$x %*% v2, col = rgb(0, 0, 0, .2),
    xlab = "optimal projection", ylab = "PC*", main = "PPCI solution",
    pch = 16)
> abline(v = pp_sol$b, col = 2, lwd = 2)
> plot(fk_density(optidigits$x %*% fk_sol$v), type = "l",
    xlab = "optimal projection", ylab = "estimated density")
> abline(v = fk_sol$b, col = 2, lwd = 2)
> plot(fk_density(optidigits$x %*% pp_sol$v), type = "l",
    xlab = "optimal projection", ylab = "estimated density")
> abline(v = pp_sol$b, col = 2, lwd = 2)
\end{verbatim}
The plots are shown in Figure~\ref{fig:optidigits_mdh}. The upper two plots show scatter plots of the data projected into two-dimensional subspaces. In each case the horizontal axis corresponds to the optimal projection obtained from MDPP, while the vertical axis is the first principal component of the data after projection into the null-space of the MDPP projection (defined in the above as \code{v2}). The lower two plots show the kernel density estimates computed from the data projected on the optimal projection vectors. It is apparent that the \mypkg~method obtained a hyperplane of considerably lower density in this example.

\begin{figure}[h]
    \centering
    \includegraphics[width = 11cm, height = 8cm]{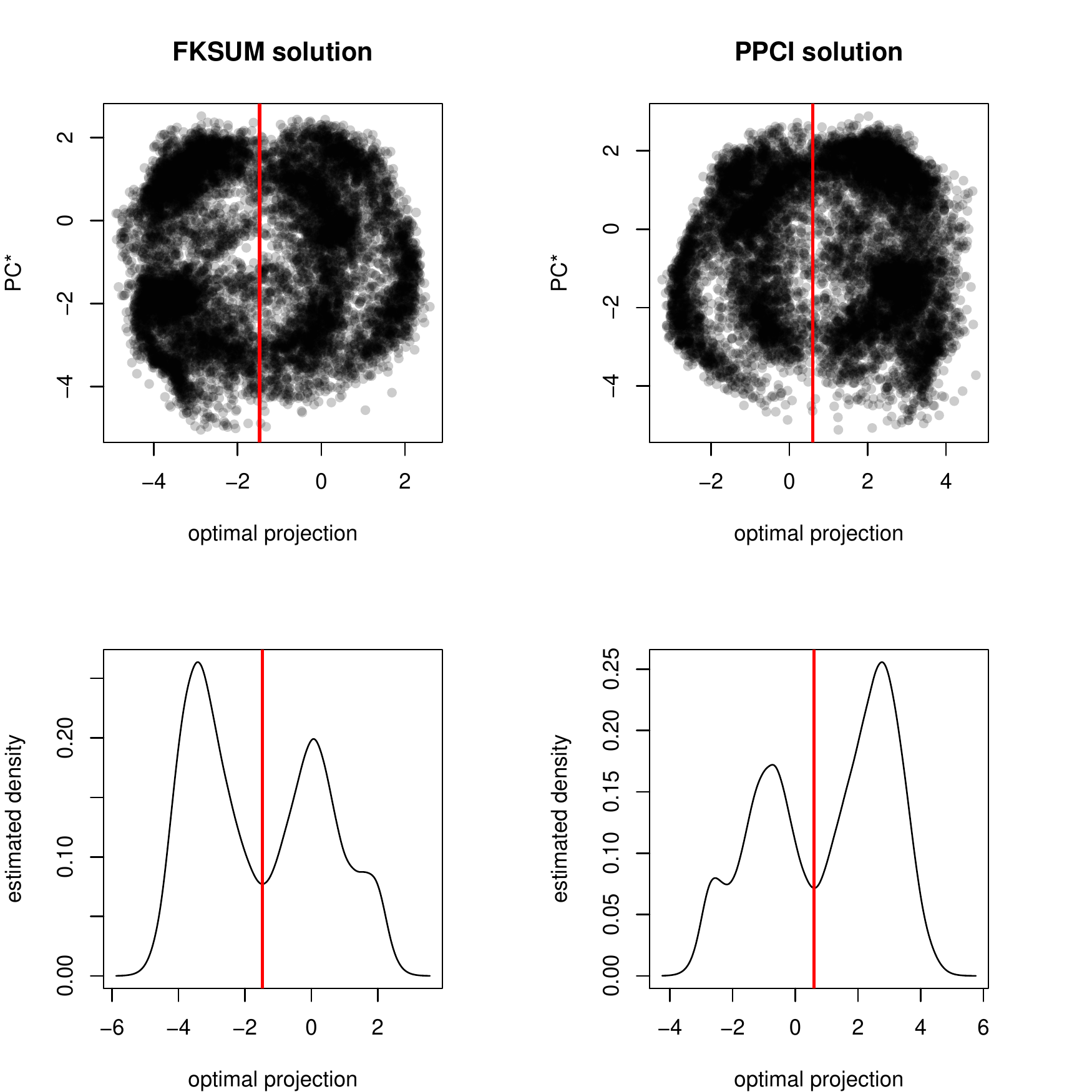}
    \caption{Minimum density hyperplane solutions on Pen-based Recognition of Handwritten Digits data. The top plots show the data projected into a two-dimensional subspace determined by the optimal projection vector and the first principal component in its null-space, while the bottom plots show the density of the data projected on the optimal projection vector.}
    \label{fig:pendigits_mdh}
\end{figure}

Next we perform the same steps for the ``Pen-based Recognition of Handwritten Digits'' data set. These data contain 16 dimensions, corresponding to variables derived from hand trajectories measured during the writing of the different digits 0--9.
\begin{verbatim}
> data("pendigits")
>
> system.time(fk_sol <- fk_mdh(pendigits$x))
   user  system elapsed 
  0.756   0.000   0.756
> success_ratio(pendigits$x %*% fk_sol$v < fk_sol$b[1], pendigits$c)
[1] 0.8477202
> system.time(pp_sol <- mdh(pendigits$x))
   user  system elapsed 
  2.545   0.021   2.567
> success_ratio(pendigits$x %*% pp_sol$v < pp_sol$b[1], pendigits$c)
[1] 0.8184627
> par(mfrow = c(2, 2))
>
> v2 <- eigen(cov(pendigits$x - 
    pendigits$x %*% fk_sol$v %*% t(fk_sol$v)))$vectors[,1]
> plot(pendigits$x %*% fk_sol$v, pendigits$x %*% v2, col = rgb(0, 0, 0, .2), 
    xlab = "optimal projection", ylab = "PC*", main = "FKSUM solution",
    pch = 16)
> abline(v = fk_sol$b, col = 2, lwd = 2)
> v2 <- eigen(cov(pendigits$x - 
    pendigits$x %*% pp_sol$v %*% t(pp_sol$v)))$vectors[,1]
> plot(pendigits$x %*% pp_sol$v, pendigits$x %*% v2, col = rgb(0, 0, 0, .2),
    xlab = "optimal projection", ylab = "PC*", main = "PPCI solution",
    pch = 16)
> abline(v = pp_sol$b, col = 2, lwd = 2)
> plot(fk_density(pendigits$x %*% fk_sol$v), type = "l",
    xlab = "optimal projection", ylab = "estimated density")
> abline(v = fk_sol$b, col = 2, lwd = 2)
> plot(fk_density(pendigits$x %*% pp_sol$v), type = "l",
    xlab = "optimal projection", ylab = "estimated density")
> abline(v = pp_sol$b, col = 2, lwd = 2)
\end{verbatim}
The implementation in \mypkg~runs considerably more efficiently and achieves a slightly better clustering accuracy. However, based on the plots in Figure~\ref{fig:pendigits_mdh}, without this {\em a posteriori} accuracy measure the two implementations appear to have obtained solutions of very similar quality in terms of the actual minimum density objective.

\subsection{Projection Pursuit Regression: A Detailed Implementation}\label{sec:ppr}

In this final discussion of projection pursuit we provide a very detailed description of the implementation of projection pursuit regression (PPR). The intention is that the reader will find sufficient instruction for the implementation of their own projection indices, or indices which are not currently offered in the \mypkg~package. This section contains considerably more mathematical detail than previous sections, and the less interested reader may wish to skip this section. For details on the use of the PPR implementation in \mypkg, use the command \code{help(fk\_ppr)}.

The standard PPR model assumes the conditional distribution of the response variable, $Y$, given a vector of covariates, $X$, is given by
\begin{align*}
Y|X \sim N\left(\mu + \sum_{j=1}^k f_j(\w_j^\top X), \sigma^2\right),
\end{align*}
where the functions $f_j:\R\to \R, j = 1, ..., k$, are assumed to come from some known class of functions, say $\C$. It is these functions, as well as the projection vectors $\w_1, ..., \w_k$ and the global mean, $\mu$, which are to be estimated. Notice that if $\C$ is the class of linear functions, then the above problem reverts to the standard linear regression model. In general $\C$ is assumed to be a very rich class of functions, where restrictions might only be placed on the number of continuous derivatives admissible by its members. In these general cases non-parametric estimation of $f_j, j = 1, ..., k$ becomes preferable. The simplest approach to fitting a PPR model follows a forward procedure in which each term in the expression for the conditional expectation of the response,
\begin{align*}
E[Y|X=\x] = \mu + \sum_{j=1}^k f_j(\w_j^\top \x),
\end{align*}
is estimated based on the residuals remaining after the terms estimated so far at each stage have been accommodated. That is, if $\{(\x_1, y_1), ..., (\x_n, y_n)\}$ represent observed pairs of the covariates and response, then initially the mean is estimated as $\hat \mu = \bar y = \frac{1}{n}\sum_{i=1}^n y_i$ and the residuals are first set to $r_i \gets y_i - \hat \mu, i = 1, ..., n$. The following is then iterated, for $j = 1, ..., k$: (i) estimate $\w_j$ and $f_j(\cdot)$ by minimising $\sum_{i=1}^n \left(r_i - f_j(\w_j^\top \x_i)\right)^2$; (ii) update the residuals by setting $r_i \gets r_i - f_j(\w_j^\top \x_i), i = 1, ..., n$. The main computational and methodological challenge, therefore, is in estimating each of the pairs $\w_j, f_j(\cdot)$, for $j = 1, ..., k$. We henceforth drop the subscript $j$ and simply focus on estimating the pair of projection vector, $\w$, and univariate regression function, $f(\cdot)$, from a set of pairs of covariates and residuals, $\{(\x_1, r_1), ..., (\x_n, r_n)\}$.

For brevity and simplicity we focus here only on using the Nadaraya-Watson estimator for $f(\cdot)$\footnote{the function \code{fk\_ppr()} in \mypkg~provides implementations of both the Nadaraya-Watson and local linear estimators.}, and for enhanced stability we will use leave-one-out estimates during optimisation. The projection index is therefore given by
\begin{align*}
    \Phi(\w|\X,\r) = \sum_{i=1}^n \left.\left(r_i - \frac{\sum_{j\not = i}K\left(\frac{p_j - p_i}{h}\right)r_j}{\sum_{j\not = i}K\left(\frac{p_j - p_i}{h}\right)}\right)^2\right|_{\p = \X\vec \w},
\end{align*}
where $\r$ is the vector of residuals and $\X$ is the matrix with $i$-th row given by $\x_i^\top$. It will be convenient to introduce the notation $\ssig(\cdot)$ and $\ssig^\prime(\cdot)$ to be the vector valued functions mapping $\R^n\to\R^n$, with
\begin{align*}
    \ssig(\om)_i &= \sum_{j\not = i} K\left(\frac{p_j - p_i}{h}\right)\omega_j,\\
    \ssig^\prime(\om)_i &= \sum_{j\not = i} K^\prime\left(\frac{p_j - p_i}{h}\right)\omega_j,
\end{align*}
where $\p = \X\vec \w$, as before, and the value of the projection vector, $\w$, will be clear from the context. For example, we may now write
\begin{align*}
    \Phi(\w|\X,\r) = \sum_{i=1}^n \left(r_i - \frac{\ssig(\r)_i}{\ssig(\one)_i}\right)^2 = \sum_{i=1}^n \left(r_i - \hat r_i\right)^2\bigg|_{\hat r_i = \frac{\ssig(\r)_i}{\ssig(\one)_i}},
\end{align*}
where $\one$ is a vector of ones, and $\hat \r = (\hat r_1, ..., \hat r_n)$ is the vector of fitted values for the residuals. We are now ready to formulate the objective function to be optimised in {R}. That is, we define the function \code{phi\_ppr()}, which evaluates the projection index for a given projection vector, as follows
\begin{verbatim}
> phi_ppr <- function(w, X, r, h, beta){
    n <- nrow(X)
    p <- X %*% w / sqrt(sum(w^2))
    Sr <- fk_sum(p, r, h, beta = beta) - beta[1] * r
    S1 <- fk_sum(p, rep(1, n), h, beta = beta) - beta[1]
    S1[S1 < 1e-20] <- 1e-20
    r_hat <- Sr / S1
    sum((r - r_hat)^2)
 }
\end{verbatim}
Here \code{Sr} and \code{S1} represent the vectors $\ssig(\r)$ and $\ssig(\one)$, respectively. As we did earlier, when using leave-one-out estimates, we buffer the sums whose evaluation as zero may cause problems, in this case because they occur in the denominator of a ratio.

Now, in order to effectively and efficiently optimise this function, we need to evaluate its gradient. 
In order to compute the gradient, we use the chain rule, as follows,
\begin{align*}
\nabla_\w \Phi(\w|\X, \r)^\top &= \nabla_{\p}\left.\left(\sum_{i=1}^n \left(r_i - \frac{\sum_{j\not = i}K\left(\frac{p_j - p_i}{h}\right)r_j}{\sum_{j\not = i}K\left(\frac{p_j - p_i}{h}\right)}\right)^2\right)^\top D_\w \p\right|_{\p = \X\vec \w},
\end{align*}
where $\nabla_\p (\cdot)$ indicates the vector of partial derivatives based on the projected sample, $\p = \X\vec\w$, and $D_\w\p$ is the matrix whose $i,j$-th entry is
the partial derivative of $p_i = \vec\w^\top \x_i$ with respect to $w_j$, which is equal to $\frac{x_{ij}}{||\w||}-\frac{p_i w_j}{||\w||^2}$. We thus have $D_\w\p = \left(\frac{1}{||\w||}\X -\frac{1}{||\w||^2}\p\w^\top\right)$. Now, if we again let $\hat r_i = \sum_{j\not=i}K\left(\frac{p_j-p_i}{h}\right)r_j/\sum_{j\not=i}K\left(\frac{p_j-p_i}{h}\right), i = 1, ..., n$, then the first term in the chain rule product can be determined using
\begin{align*}
     \frac{\partial}{\partial p_i}\sum\limits_{j=1}^n \left(r_j - \hat r_j\right)^2 &= \sum\limits_{j=1}^n \frac{\partial}{\partial \hat r_j}\left(r_j - \hat r_j\right)^2 \frac{\partial \hat r_j}{\partial p_i}\\
    &= 2\sum\limits_{j=1}^n \left(\hat r_j - r_j\right) \frac{\partial }{\partial p_i} \frac{\sum\limits_{k\not =j}^n K\left(\frac{p_k - p_j}{h}\right)r_k}{\sum\limits_{k\not =j}^nK\left(\frac{p_k - p_j}{h}\right)}\\
    &= 2\sum_{j=1}^n \left(\hat r_j - r_j\right)\left( \frac{\sum\limits_{k\not =j}^ny_k \frac{\partial }{\partial p_i} K\left(\frac{p_k - p_j}{h}\right)}{\sum\limits_{k\not =j}^nK\left(\frac{p_k - p_j}{h}\right)} - \frac{\sum\limits_{k\not =j}^n K\left(\frac{p_k - p_j}{h}\right)r_k\sum\limits_{k\not =j}^n\frac{\partial }{\partial p_i}K\left(\frac{p_k - p_j}{h}\right)}{\left(\sum\limits_{k\not =j}^nK\left(\frac{p_k - p_j}{h}\right)\right)^2}\right)\\
    &= 2\sum\limits_{j=1}^n \frac{\hat r_j - r_j}{\sum\limits_{k\not =j}^n K\left(\frac{p_k - p_j}{h}\right)}\left(\sum\limits_{k\not =j}^nr_k \frac{\partial }{\partial p_i} K\left(\frac{p_k - p_j}{h}\right) - 
    \hat r_j\sum\limits_{k\not =j}^n\frac{\partial }{\partial p_i}K\left(\frac{p_k - p_j}{h}\right)\right)\\
    &= 2\sum\limits_{j\not = i} \frac{\hat r_j - r_j}{\sum\limits_{k\not =j}^n K\left(\frac{p_k - p_j}{h}\right)}\left(\sum\limits_{k\not =j}^nr_k \frac{\partial }{\partial p_i} K\left(\frac{p_k - p_j}{h}\right) - 
    \hat r_j\sum\limits_{k\not =j}^n\frac{\partial }{\partial p_i}K\left(\frac{p_k - p_j}{h}\right)\right)\\
    & \ \ \ \ + 2\frac{\hat r_i - r_i}{\sum\limits_{k\not = i}^n K\left(\frac{p_k - p_i}{h}\right)}\left(\sum\limits_{k\not =i}^nr_k \frac{\partial }{\partial p_i} K\left(\frac{p_k - p_i}{h}\right) - 
    \hat r_i\sum\limits_{k\not =i}^n\frac{\partial }{\partial p_i}K\left(\frac{p_k - p_i}{h}\right)\right)\\
    &= 2\sum\limits_{j\not = i} \frac{\hat r_j - r_j}{\sum\limits_{k\not =j}^n K\left(\frac{p_k - p_j}{h}\right)}\left( \frac{r_i}{h} K'\left(\frac{p_i - p_j}{h}\right) - 
    \frac{\hat r_j}{h}K'\left(\frac{p_i - p_j}{h}\right)\right)\\
    & \ \ \ \ - 2\frac{\hat r_i - r_i}{\sum\limits_{k\not =i}^n K\left(\frac{p_k - p_i}{h}\right)}\left(\frac{1}{h}\sum\limits_{k\not =i}^nr_k K'\left(\frac{p_k - p_i}{h}\right) - 
    \frac{\hat r_i}{h}\sum\limits_{k\not =i}^nK'\left(\frac{p_k - p_i}{h}\right)\right)\\
    &= \frac{2}{h}\sum\limits_{j\not=i}\frac{\hat r_j(\hat r_j-r_j)}{\sum\limits_{k\not = j}K\left(\frac{p_k-p_j}{h}\right)}K'\left(\frac{p_j-p_i}{h}\right)-\frac{2r_i}{h}\sum\limits_{j\not=i}\frac{(\hat r_j-r_j)}{\sum\limits_{k\not = j}K\left(\frac{p_k-p_j}{h}\right)}K'\left(\frac{p_j-p_i}{h}\right)\\
    & \ \ \ \ + \frac{2\hat r_i(\hat r_i-r_i)}{h\sum\limits_{k\not=i}K\left(\frac{p_k-p_i}{h}\right)}\sum_{j\not=i}K'\left(\frac{p_j-p_i}{h}\right)- \frac{2(\hat r_i-r_i)}{h\sum\limits_{k\not=i}K\left(\frac{p_k-p_i}{h}\right)}\sum\limits_{j\not=i}r_jK'\left(\frac{p_j-p_i}{h}\right),
\end{align*}
where some of the signs have changed in the final step due to the reversing of the terms inside $K'(\cdot)$, which is an odd function, i.e., $K'(x) = -K'(-x) \ \forall x$. This is now in a form convenient for the notation $\ssig(\cdot), \ssig'(\cdot)$ used previously. That is, we find
\begin{align*}
    \frac{\partial}{\partial p_i} \sum_{j=1}^n(r_j - \hat r_j) &= \frac{2}{h}\left(\ssig'\left(\frac{\hat \r(\hat \r-\r)}{\ssig(\one)}\right)_i-r_i\ssig'\left(\frac{\hat \r-\r}{\ssig(\one)}\right)_i + \frac{\hat r_i-r_i}{\ssig(\one)_i}\left(\hat r_i\ssig'(\one)_i-\ssig'(\r)_i\right)\right),
\end{align*}
where products and ratios of vectors within $\ssig'(\cdot)$ are element-wise operations. To evaluate the gradient, both the kernel and kernel derivative sums of $\r$ and $\one$ are required. We can therefore set \code{type = "both"} in the function \code{fk\_sum()}. For the other sums, we only require the sums of kernel derivatives. To compute the gradient of the projection index in {R}, we can thus use the following,
\begin{verbatim}
> dphi_ppr <- function(w, X, r, h, beta){
    n <- nrow(X)
    nw <- sqrt(sum(w^2))
    p <- X %*% w / nw
    S1 <- fk_sum(p, rep(1, n), h, beta = beta, type = "both")
    S1[, 1] <- S1[, 1] - beta[1]
    S1[S1[, 1] < 1e-20, 1] <- 1e-20
    Sr <- fk_sum(p, r, h, beta = beta, type = "both")
    Sr[, 1] <- Sr[, 1] - beta[1] * r
    r_hat <- Sr[, 1] / S1[, 1]
    
    T1 <- fk_sum(p, r_hat * (r_hat - r) / S1[, 1], h,
        beta = beta, type = "dksum")
    T2 <- r * fk_sum(p, (r_hat - r) / S1[, 1], h,
        beta = beta, type = "dksum")
    T3 <- (r_hat - r) / S1[, 1] * (r_hat * S1[, 2] - Sr[, 2])
    dphi_dp <- (T1 - T2 + T3) * 2 / h
    dp_dw <- (X / nw - p %*% t(w) / nw^2)
    c(t(dphi_dp) %*% dp_dw)
 }
\end{verbatim}
In the above we have split the partial derivatives with respect to the elements in $\p$ into terms \code{T1, T2, T3}.
We have also used the fact that the function \code{fk\_sum(..., type = "both")} returns a matrix in which the first column contains the kernel sums and the second contains the sums of kernel derivatives. Note that since $K'(0) = 0$ for all symmetric, differentiable kernels, the sums of kernel derivatives are the same whether they are of the leave-one-out type or not. Hence, only the first column in the output of \code{fk\_sum(..., type = "both")} needs to be modified to accommodate the fact that we use leave-one-out sums.

Before continuing, it is prudent to verify that the gradient function is producing the correct output. In order to do so, we compare its output with a numerically approximated gradient based on finite differences. In particular, if a function, $g(\cdot)$, is differentiable, then,
\begin{align*}
    \frac{\partial}{\partial w_i} g(\w) = \frac{g(\w + h\mathbf{e}_i)-g(\w - h\mathbf{e}_i)}{2h} + o(h),
\end{align*}
where $\mathbf{e}_i$ is the vector of zeroes except in the $i$-th position, where it takes the value one.
%
Note that a simple way to encode the collection of vectors $\{\mathbf{e}_1, ..., \mathbf{e}_d\}$ is as the rows of the identity matrix. We begin by generating a set of data (covariates and response variable). We will generate 1000 data points in 10 dimensions from a multivariate Gaussian distribution with a randomly generated covariance matrix. We will then select some random ``true'' projection vectors, and define the response as a non-linear function of the projected data.
\begin{verbatim}
> set.seed(1)
> n_dat <- 1000
> n_dim <- 10
>
> X <- matrix(rnorm(n_dat * n_dim), n_dat, n_dim) %*%
    matrix(2 * runif(n_dim^2) - 1, n_dim, n_dim)
>
> wtrue1 <- rnorm(n_dim)
> wtrue2 <- rnorm(n_dim)
> 
> y <- (X %*% wtrue1 > 1) * (X %*% wtrue1 - 1) + tanh(X %*% wtrue2 / 2) * 
    (X %*% wtrue1) + (X %*% (wtrue1 - wtrue2) / 5)^2 + rnorm(n_dat)
\end{verbatim}
We now check that our exact gradient matches closely to the finite differences approximation, both in a relative and absolute sense, for a randomly selected projection vector. We also, for now, simply choose a random positive bandwidth value.
\begin{verbatim}
> w <- rnorm(n_dim)
> h <- runif(1)
> beta <- c(0.25, 0.25)
>
> Eh <- diag(n_dim) * 1e-5
> dphi_approx <- apply(Eh, 1, function(eh) (phi_ppr(w + eh, X, y, h, beta)
    - phi_ppr(w - eh, X, y, h, beta)) / 2e-5)
> dphi <- dphi_ppr(w, X, y, h, beta)
> max(abs(dphi / dphi_approx - 1))
[1] 7.004954e-10
> max(abs(dphi - dphi_approx))
[1] 1.47976e-07
\end{verbatim}
We see that the analytical and numerically approximated gradients are extremely close to one another, and so should feel confident that our analytical expression, and the function for evaluating it, is correct. 

\begin{figure}
    \centering
    \includegraphics[width = \textwidth]{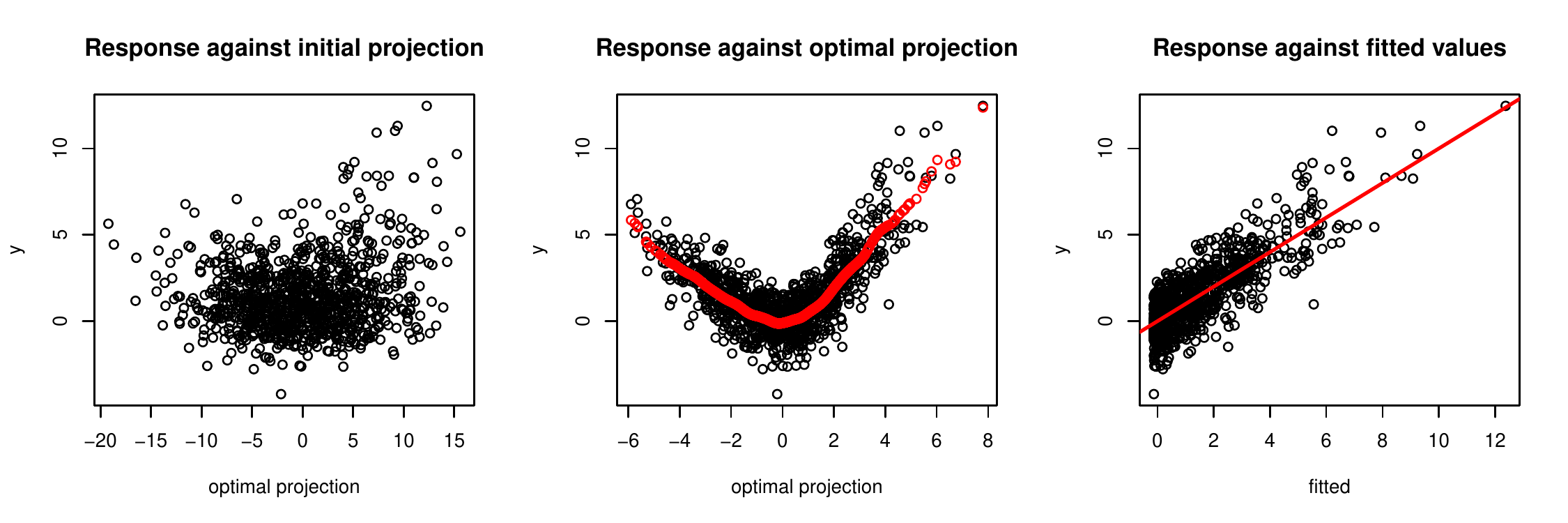}
    \caption{Projection pursuit regression solution from ten dimensional simulated data. The plots show the response against the initial projection, the optimal projection and the fitted values.}
    \label{fig:ppr1}
\end{figure}

Next we write a function, \code{ppr\_nw()}, to perform projection pursuit based on minimising \code{phi\_ppr()}. To optimise the projection we apply the function \code{optim()}, from {R}'s base \pkg{stats} package. Our preferred optimisation method is the limited memory BFGS algorithm~\citep{BFGS}. We have found that using an oversmoothing bandwidth during projection pursuit is quite reliable, and tends to be fairly robust as the estimation along each projection has relatively low variation. We select this bandwidth using the optimal rate for univariate regression, $\mathcal{O}(n^{-1/5})$, and a measure of the scale of the covariates, for which we use the square root of the largest eigenvalue of their covariance matrix. Once the optimal projection has been determined, however, the final fitting is performed using a more sophisticated method which is based on leave-one-out cross-validation. It is generally preferable to initialise the projection vector, $\w$, with a sensible heuristic. We use the ordinary least squares solution, with a small ridge to ensure a unique solution, $\w = (\X^\top \X + 0.01\mathbf{I})^{-1}\X^\top \y$. We also allow for alternative initialisation with the optional argument \code{w}, which is by default set to this ridge solution.
\begin{verbatim}
> ppr_nw <- function(X, r, w = NULL){
    n <- nrow(X)
    d <- ncol(X)
    if(is.null(w)) w <- solve(t(X) %*% X + .01 * diag(d)) %*% t(X) %*% r
    h <- sqrt(eigen(cov(X))$values[1]) / n^.2
    w <- optim(w, phi_ppr, dphi_ppr, X, r, h, c(.25, .25), 
        method = "L-BFGS-B")$par
    w <- w / sqrt(sum(w^2))
    loo_sse <- function(h) phi_ppr(w, X, r, h, c(.25, .25))
    h <- optimise(loo_sse, c(h/50, h))$minimum
    list(w = w, h = h)
 }
\end{verbatim}
We now find the optimal projection for the data which we generated previously, and plot the result. For interest we also plot the response values against the projections along the initial randomly generated projection, as well as against the fitted values. The plots can be seen in Figure~\ref{fig:ppr1}. Notice that the implementation we have provided above does not necessitate that the mean of the response needs to first be subtracted to obtain the first set of remaining residuals. However, to be consistent with the discussion above, we will include this step.
\begin{verbatim}
> mu <- mean(y)
> r <- y - mu
>
> w_opt <- ppr_nw(X, r, w = w)
>
> p <- X %*% w_opt$w
> S1 <- fk_sum(p, rep(1, n_dat), w_opt$h)
> Sr <- fk_sum(p, r, w_opt$h)
> fitted <- mu + Sr / S1
>
> par(mfrow = c(1, 3))
> 
> plot(X %*% w, y, main = "Response against initial projection", 
    xlab = "optimal projection", ylab = "y")
> plot(X %*% w_opt$w, y, main = "Response against optimal projection",
    xlab = "optimal projection", ylab = "y")
> points(X %*% w_opt$w, fitted, col = 2)
> plot(fitted, y, main = "Response against fitted values",
    xlab = "fitted", ylab = "y")
> abline(0, 1, lwd = 2, col = 2)
\end{verbatim}

\paragraph{Example: Simulation} Once again we begin with a simulation to assess the speed and accuracy of our implementation. We will compare with the function \code{ppr()} from {R}'s \pkg{stats} package. This function uses an iterative re-weighted least squares approach to optimisation, as opposed to a more standard gradient descent. As a result, this implementation is extremely efficient for problems of moderate dimensionality, but scales quadratically in the problem dimension\footnote{Many gradient descent methods scale linearly in the problem dimension.}. We therefore consider two scenarios, one of low dimensionality and another of higher dimensionality. We simulate pairs of response and covariates exactly as we did above. We then use half of the data for training, and the other half to estimate the coefficient of determination of the models. We repeat this experiment 50 times.
\begin{verbatim}
> n_rep <- 50
>
> t_stats <- numeric(n_rep)
> t_nw <- numeric(n_rep)
> R2_stats <- numeric(n_rep)
> R2_nw <- numeric(n_rep)
>
> for(rep in 1:n_rep){
    set.seed(rep)
    X <- matrix(rnorm(n_dat * n_dim), n_dat, n_dim) %*%
        matrix(2 * runif(n_dim * n_dim) - 1, n_dim, n_dim)
    wtrue1 <- rnorm(n_dim)
    wtrue2 <- rnorm(n_dim)
    y <- (X %*% wtrue1 > 1) * (X %*% wtrue1 - 1) + tanh(X %*% wtrue2 / 2) *
        (X %*% wtrue1) + (X %*% (wtrue1 - wtrue2) / 5)^2 + rnorm(n_dat)
        
    t_stats[rep] <- system.time(model <- ppr(X[1:(n_dat / 2),],
        y[1:(n_dat / 2)], nterms = 1))[1]
    yhat <- predict(model, X[(n_dat / 2 + 1):n_dat,])
    R2_stats[rep] <- 1 - mean((yhat - y[(n_dat / 2 + 1):n_dat])^2) / 
        var(y[(n_dat / 2 + 1):n_dat])
        
    t_nw[rep] <- system.time(model <- ppr_nw(X[1:(n_dat / 2),], 
        y[1:(n_dat / 2)] - mean(y[1:(n_dat / 2)])))[1]
    p <- X[1:(n_dat / 2),] %*% model$w
    ptest <- X[(n_dat / 2 + 1):n_dat,] %*% model$w
    S1 <- fk_sum(p, rep(1, n_dat / 2), model$h, x_eval = ptest)
    Sr <- fk_sum(p, y[1:(n_dat / 2)] - mean(y[1:(n_dat / 2)]), model$h, 
    	x_eval = ptest)
    yhat <- mean(y[1:(n_dat / 2)]) + Sr / S1
    R2_nw[rep] <- 1 - mean((yhat - y[(n_dat / 2 + 1):n_dat])^2) / 
        var(y[(n_dat / 2 + 1):n_dat])
 }
>
> colMeans(cbind(t_stats, t_nw))
t_stats    t_nw 
0.00299 0.03372
> colMeans(cbind(R2_stats, R2_nw))
 R2_stats     R2_nw 
0.6948070 0.6493932
\end{verbatim}
In this experiment the existing implementation is superior in both speed and accuracy. Next we consider a much higher dimensional example, containing 200 covariates. We also increase the number of data to 5000. In the interest of time, we repeat this experiment only 20 times, and again report the results.
\begin{verbatim}
> n_dat <- 5000
> n_dim <- 200
>
> n_rep <- 20
>
> t_stats <- numeric(n_rep)
> t_nw <- numeric(n_rep)
> R2_stats <- numeric(n_rep)
> R2_nw <- numeric(n_rep)
>
> for(rep in 1:n_rep){
    set.seed(rep)
    X <- matrix(rnorm(n_dat * n_dim), n_dat, n_dim) %*%
        matrix(2 * runif(n_dim * n_dim) - 1, n_dim, n_dim)
    wtrue1 <- rnorm(n_dim)
    wtrue2 <- rnorm(n_dim)
    y <- (X %*% wtrue1 > 1) * (X %*% wtrue1 - 1) + tanh(X %*% wtrue2 / 2) *
        (X %*% wtrue1) + (X %*% (wtrue1 - wtrue2) / 5)^2 + rnorm(n_dat)
        
    t_stats[rep] <- system.time(model <- ppr(X[1:(n_dat / 2),],
        y[1:(n_dat / 2)], nterms = 1))[1]
    yhat <- predict(model, X[(n_dat / 2 + 1):n_dat,])
    R2_stats[rep] <- 1 - mean((yhat - y[(n_dat / 2 + 1):n_dat])^2) / 
        var(y[(n_dat / 2 + 1):n_dat])
        
    t_nw[rep] <- system.time(model <- ppr_nw(X[1:(n_dat / 2),],
        y[1:(n_dat / 2)] - mean(y[1:(n_dat / 2)])))[1]
    p <- X[1:(n_dat / 2),] %*% model$w
    ptest <- X[(n_dat / 2 + 1):n_dat,] %*% model$w
    S1 <- fk_sum(p, rep(1, n_dat / 2), model$h, x_eval = ptest)
    Sr <- fk_sum(p, y[1:(n_dat / 2)] - mean(y[1:(n_dat / 2)]), model$h,
    	 x_eval = ptest)
    yhat <- mean(y[1:(n_dat / 2)]) + Sr / S1
    R2_nw[rep] <- 1 - mean((yhat - y[(n_dat / 2 + 1):n_dat])^2) / 
        var(y[(n_dat / 2 + 1):n_dat])
 }
>
> colMeans(cbind(t_stats, t_nw))
t_stats    t_nw 
 3.2234  1.1164 
> colMeans(cbind(R2_stats, R2_nw))
  R2_stats      R2_nw 
-0.7908563  0.7890950
\end{verbatim}
In this case the existing implementation overfits the data, and thus obtains a large negative coefficient of determination. On the other hand, our implementation continues to provide accurate predictions. Also, as expected, the running time of the existing implementation increases substantially, now exceeding that of ours.

\paragraph{Example: Baseball Salaries} To conclude this section we consider a simple real data example taken from \cite{ISLR}, which is available in the package \pkg{ISLR}~\citep{ISLRCRAN}. The problem is to predict the salaries of professional baseball players based on various performance statistics. For simplicity, we only use the numerical covariates. We begin, as always, by loading the required package.
\begin{verbatim}
> if(! "ISLR" %in% installed.packages()) install.packages("ISLR")
> library(ISLR)
\end{verbatim}
Next we set up the matrix of covariates and the vector of responses, removing those cases which do not have a response value.
\begin{verbatim}
> X <- as.matrix(Hitters[! is.na(Hitters[, 19]), c(1:13, 16:18)])
> y <- Hitters[! is.na(Hitters[, 19]), 19]
\end{verbatim}
We will again compare models based on their estimated coefficient of determination. Because the data set is small, containing only 263 complete observations, we will consider multiple splits into 70\% training and 30\% testing data. We store the training indices in the vector \code{train}. For interest we also consider models with two components. To estimate the second component, we simply apply the function \code{ppr\_nw()} to the data using the residuals after the first component's fitted values have also been subtracted from the responses. 
\begin{verbatim}
> n_rep <- 50
>
> R2_stats_1 <- numeric(n_rep)
> R2_stats_2 <- numeric(n_rep)
> R2_nw_1 <- numeric(n_rep)
> R2_nw_2 <- numeric(n_rep)
>
> for(rep in 1:n_rep){
    set.seed(rep)
    train <- sample(1:nrow(X), floor(.7 * nrow(X)))
    n_train <- length(train)
    
    model <- ppr(X[train,], y[train], nterms = 1)
    yhat <- predict(model, X[-train,])
    R2_stats_1[rep] <- 1 - mean((yhat - y[-train])^2) / var(y[-train])
    
    model <- ppr(X[train,], y[train], nterms = 2)
    yhat <- predict(model, X[-train,])
    R2_stats_2[rep] <- 1 - mean((yhat - y[-train])^2) / var(y[-train])
    
    mu <- mean(y[train])
    r <- y[train] - mu
    component_1 <- ppr_nw(X[train,], r)
    p <- X %*% component_1$w
    S1 <- fk_sum(p[train], rep(1, n_train), component_1$h, x_eval = p)
    Sr <- fk_sum(p[train], r, component_1$h, x_eval = p)
    fitted <- Sr[train] / S1[train]
    yhat <- mu + Sr[-train] / S1[-train]
    R2_nw_1[rep] <- 1 - mean((yhat - y[-train])^2) / var(y[-train])
    
    r <- r - fitted
    component_2 <- ppr_nw(X[train,], r)
    p <- X %*% component_2$w
    S1 <- fk_sum(p[train], rep(1, n_train), component_2$h, x_eval = p[-train])
    Sr <- fk_sum(p[train], r, component_2$h, x_eval = p[-train])
    
    yhat <- yhat + Sr / S1
    R2_nw_2[rep] <- 1 - mean((yhat - y[-train])^2) / var(y[-train])
 }
>
> colMeans(cbind(R2_stats_1, R2_stats_2, R2_nw_1, R2_nw_2))
R2_stats_1 R2_stats_2     R2_nw_1    R2_nw_2 
 0.2985216  0.1693295  0.3568934  0.4341185  
\end{verbatim}
Our implementation not only obtains superior accuracy to the existing implementation, but also better utilises the added flexibility of the additional component, whereas this flexibility appears to lead to overfitting when using the function \code{ppr()} in this instance.

\section{Conclusion}\label{sec:conclusion}

In this paper we discussed the use of the {R} package \mypkg~for efficient univariate kernel smoothing and projection pursuit when the projection index requires evaluating kernel based estimates of functionals of the projected data distribution. These include independent component analysis; projection pursuit regression and minimum density hyperplanes. In all cases we investigated the accuracy and efficiency of the package's implementation in comparison with existing implementations available either with {R}'s standard distribution or through the comprehensive {R} archive network. In all cases the implementations provided in \mypkg~showed very competitive performance. In addition we provided a detailed derivation and implementation of projection pursuit regression, which we believe should provide readers with sufficient instruction to use the package functionality for implementing smoothing and projection pursuit methods in {R} which lie beyond the scope of the package in its current form. 

\bibliographystyle{plainnat}
\bibliography{main}


\end{document}